\documentclass[12pt,a4paper]{article}
\title{Optimal Block Designs}
\author{Eric Nyarko}
\title{\textbf{Technical Report: Optimal Paired Comparison Block Designs}}
\author{Eric Nyarko\footnote{corresponding author.~E-mail:~\texttt{eric.nyarko@ovgu.de; nyarkoeric5@gmail.com}}\\%, Rainer Schwabe~\footnote{E-mail:~\texttt{rainer.schwabe@ovgu.de}}\\
University of Magdeburg, \\
Institute for Mathematical Stochastics,\\
PF 4120, D-39016 Magdeburg, Germany}
\usepackage{amsmath}
\usepackage{booktabs}
\usepackage{dirtytalk}
\usepackage{float}
\usepackage{float,graphicx}
\usepackage[figuresright]{rotating}
\usepackage{verbatim}
\usepackage{diagbox}
\usepackage{slashbox}
\usepackage{setspace}
\usepackage{appendix}
\usepackage{tikz}
\usepackage{longtable,pdflscape,caption,hyperref}
\usepackage{amsmath,mathtools}
\usepackage{amsthm}
\usepackage{siunitx}
\usepackage{enumitem}
\usepackage{enumerate}
\usepackage{graphicx}
\usepackage{makecell}

\usepackage{amsmath,amssymb,amsfonts,mathrsfs,microtype, amsthm}
\usepackage[flushleft]{threeparttable}

\usepackage{natbib}
\usepackage{lipsum} 
\usepackage{dcolumn}     
\date{}

\begin{document}
\maketitle
\begin{abstract}\noindent
For paired comparison experiments involving competing options described by two-level attributes several different methods of constructing designs having block paired observations under the main effects model are presented. These designs are compared to alternative designs available in the literature.
\end{abstract}

\section{Preliminaries} \label{Chapter5} % For referencing the chapter elsewhere, use \ref{Chapter1} 
Paired comparison is a commonly used method to elicit the preference behavior of consumers and to determine the benefits that the various attributes of a product have. The pairs presented to consumers require the comparison of product descriptions compiled using an experimental design. The quality of the results of such an experiment depends on the design used. However, when choosing the design as well as modeling the data, practitioners or reserachers often fail to take into account that the respondents are asked multiple questions and that the resulting answers may therefore be correlated. \par
This technical report addresses this issue by presenting several different methods of constructing experimental designs having block paired observations for main effects within  available classes of designs. Within such classes of designs several results are known \citep[][amongst others]{jacroux1983optimality,cheng1978optimality,mukerjee2002optimal,saharay2018optimality}. Our designs are compared to that of \citet{singh2015optimal} in Table \ref{tab4.9} under the main effects block model.\par

In what follows, we define by $N=bm$ the number of paired comparisons having $b$ blocks of size $m$ and restrict our attention to the linear paired comparison model with block effects 
\begin{equation}\label{eq:4}
\textbf{Y}=  (\textbf{F}_1-\textbf{F}_2) \boldsymbol{\beta}+\textbf{Z}\boldsymbol{\gamma}+ \boldsymbol{\epsilon}=  \textbf{F} \boldsymbol{\beta}+\textbf{Z}\boldsymbol{\gamma}+ \boldsymbol{\epsilon},
\end{equation}
where $\textbf{Y}$ is a vector of $N$ pairs of responses with difference matrix $\textbf{F}=\textbf{F}_1-\textbf{F}_2$. Both $\textbf{Z} = \textbf{I}_{b}\otimes \textbf{1}_{m}$ and $\boldsymbol{\gamma}= (\boldsymbol{\gamma}_{1},\dots,\boldsymbol{\gamma}_{b})^{\top}$ denote the block indicator matrix and a vector of $b$ block effects, respectively. Moreover, $\boldsymbol{\beta}= (\beta_{1},\dots,\beta_{p})^{\top}$ and $\boldsymbol{\epsilon}$ are already defined.  \par

We now consider the situation where the paired comparison model with fixed block effects \eqref{eq:4} is to be used to study a couple of, say, $K$ attributes $k=1,\dots,K$ each at two-levels $(v_k=2)$. As already mentioned, here the experimental situation involve $N$ pairs which can be generally partitioned into $b$ blocks of size $m$. Here we allocate $N=bm$ pairs among $b$ respondents in such a way that each respondent observes $m$ pairs \citep[e.g., see][]{singh2015optimal}. Let $\Xi_N$ be the class of all such available paired comparion block designs. Under model \eqref{eq:4} and for a given paired comparion block design $\bar{\xi}\in\Xi_N$ having $N$ pairs and $K$ attributes, the least square estimate for the main effects parameter $\boldsymbol{\beta}$ is any solution to the reduced normal equation for main effects given by
\begin{equation*}
\mathbf{M}(\bar{\xi})\hat{\boldsymbol{\beta}}= \mathbf{Q},
\end{equation*}
 where
\begin{equation}\label{eq:4.47}
\mathbf{M}(\bar{\xi})= \frac{1}{4N}\Big(\mathbf{F}^{\top}\mathbf{F}-\frac{1}{m}\mathbf{F}^{\top}\mathbf{Z}\mathbf{Z}^{\top}\mathbf{F} \Big),
\end{equation}
and $\mathbf{Q}= \frac{1}{4N}\Big(\mathbf{F}^{\top}\left(\begin{matrix}\mathbf{I}_N-\frac{1}{m}\mathbf{Z}\mathbf{Z}^{\top}\end{matrix}\right)\mathbf{Y}\Big)$ where $\mathbf{I}$ denotes identity matrix of order $q$, $\mathbf{Z} = \mathbf{I}_{b}\otimes \mathbf{1}_{m}$ is the block indicator matrix of dimension $bm\times b$ and $\mathbf{F}$ is the design or difference matrix corresponding to the main effect minimal vector parameters $\boldsymbol{\beta}$ of interest of the Gauss-Markov estimator $\hat{\boldsymbol{\beta}}$ having covariance matrix of the form $\sigma^2(\textbf{F}^{\top}\textbf{F})^{-1}$ whenever $\mathbf{M}(\bar{\xi})$ is nonsingular. We note that the design $\bar{\xi}\in\Xi_N$ can be orthogonally blocked into $b$ blocks of size $m$ when the rows of the design matrix $\mathbf{F}$ are assigned to blocks as specified by $\mathbf{Z} = \mathbf{I}_{b}\otimes \mathbf{1}_{m}$. In particular, $\mathbf{F}^{\top}\mathbf{Z}=\mathbf{0}$ and the information matrix $\mathbf{M}(\bar{\xi})$ of the paired comparion block design $\bar{\xi}$ of the main effect minimal vector parameters $\boldsymbol{\beta}$ takes the form 
\begin{equation}\label{eq:4.48}
\mathbf{M}(\bar{\xi})= \frac{1}{4N}\mathbf{F}^{\top}\mathbf{F}, 
\end{equation}
which is a desirable condition in practice.  \par
%\left(\begin{smallmatrix}K-1 \\ S-1\end{smallmatrix}\right)
We mention that under the indifference assumption of equal utilities the designs considered in this paper carry over to the \citet{bradley1952rank} type choice experiments \citep[see][]{street2007construction}. In particular, this assumption simplifies the information matrix of the binary logit model which becomes proportional to the information matrix in the linear paired comparison model. This is the approach taken by \citet{grossmann2015handbook}, amongs other works.
\section{Construction Methods}

The underlying methods of construction rely on the well-known Hadamard matrix. We define a square matrix with elements $\pm1$ and size $n$ whose distinct row vectors are mutually orthogonal as a Hadamard matrix, say, $\mathbf{H}_n$ of order $n$. The smallest examples of a Hadamard matrix $\mathbf{H}_n$ of order $n$ are 
\begin{equation*}
\begin{pmatrix}1\end{pmatrix}, \begin{pmatrix}1& 1\\ 1 & -\\\end{pmatrix}, \begin{pmatrix}1& 1 & 1 & 1 \\ 1 &- & - &1\\1 & 1 & - & - \\1    & - & 1 & -\\\end{pmatrix}
\end{equation*} 
where $-$ denotes $-1$. For convenience in notations, we note throughout the sequel that the dimension of any given Hadamard matrix $\mathbf{H}_n$ is either $1$, $2$ or a multiple of $4$. For the existance of Hadamard matrices of multiples of $4$ up to $764$ \citep[e.g., see][]{djokovic2008hadamard}.\par
Further throughout the sequel we denote by $\Xi_{(N,K)}$ a class of two-level main effects paired comparison designs having attributes $K$  occuring in $N$ pairs and denote by $\Xi_{(N,K,b)}$ the class of two-level main effects paired comparison designs available when $K$ attributes occur in $N$ pairs arranged in $b$ blocks of size $m=N/b$. It immediately follows from Theorem $1$ of \citet{grasshoff2004optimal} that by using the single-attribute, main effects optimal paired comparison block designs $\bar{\xi^{\ast}}$ within the corresponding classes $\Xi_{(N,K,b)}$ can be constructed for fixed number of attributes, $k=1,\dots,K$. In particular, for the single-attribute at levels $v=2$, we let $\mathbf{A}$ be the value from the corresponding exact optimal design $\bar{\xi}^{\ast}_{v,1}$ and let $\mathbf{A}_{v}$ be the column vector corresponding to the exact optimal design $\bar{\xi}^{\ast}_{v,0}$ where $\mathbf{1}_{m}^{\top}\mathbf{A}_{v}=\mathbf{0}$ satisfying the column vector $\mathbf{1}_{m}$ of ones \citep[see][p.~364]{grasshoff2004optimal}. We further note that the symbol \say{$\otimes$} is the usual Kronecker product of vectors and matrices, and $\mathbf{J}_{rs}$ denotes a  matrix of dimension $r\times s$ of all elements equal to $1$.
\par

We now demonstrate different construction methods of the main effects optimal paired comparison block designs $\bar{\xi^{\ast}}$ within a given classes. Examples of such paired comparison block designs are exhibited in Table \ref{tab4.9} for fixed number of attributes with given number of blocks and block sizes. For notational convenience, we shall generally use a decorated $d$ as a design within a given class to distinguish between designs under the various steps of the methods of construction.

 \newpage
\section*{Method~1: }\label{mth1}%$n\equiv2$ $(\bmod \ 4)$}
For this method of constructing main effects paired comparison block designs $\bar{\xi^{\ast}}$, let $\Xi_{(N,K,b)}$ be a class of designs for $\bar{\xi^{\ast}}\in\Xi_{(N,K,b)}$ where $N\equiv2$ $(\bmod \ 4)$ and block size $m=N/b$ is a multiple of $2$. Now let $d\in\Xi_{(N/2-1,K-1)}$ denote a main effects design having matrix $\textbf{F}_{\hat{d}}$ with components of $\mathbf{A}$ such that $\textbf{F}^{\top}_{\hat{d}}\textbf{F}_{\hat{d}}=(N/2-1)\textbf{I}_{K-1}$ where $\textbf{F}_{\hat{d}}$ corresponds to $K-1$ columns of an $(n/2-1)\times(n/2-1)$ Hadamard matrix.
%For this case of design construction, we remark that the desired design $d^*$ is optimal by $\textbf{Theorem 3.3}$ \citep[e.g., see][p.~1026]{jacroux2013optimal}. 
Now the design is constructed as follows:
\newlist{steps}{enumerate}{1}
\setlist[steps, 1]{label = \textbf{Step} \arabic*:}
\begin{steps}
 \item Add a row of all $2$'s or all $-2$'s to $\textbf{F}_{\hat{d}}$  to obtain $\tilde{d}\in \Xi_{(N/2-1,K-1)}$ having
\begin{equation*}
\textbf{F}^{\top}_{\tilde{d}}\textbf{F}_{\tilde{d}}=(N/2-1)\textbf{I}_{K}+\textbf{J}_{K-1}
\end{equation*}
\item Obtain $\bar{d}$ having matrix
\begin{equation*}
\textbf{F}_{\bar{d}}=(\mathbf{A}_{N/2},\textbf{F}_{\tilde{d}}),
\end{equation*}
where $\mathbf{A}_{N/2}$ is the corresponding value $\mathbf{A}$ of dimension $N/2$, and observe that $\bar{d}\in\Xi_{(N,K,N/2)}$ where blocks consist of successive two profiles in $\textbf{F}_{\bar{d}}$. Also observe that 
\begin{equation*}
\textbf{F}^{\top}_{\bar{d}}\textbf{F}_{\bar{d}}=(N-2)\textbf{I}_{K}+2\textbf{J}_{K}.
\end{equation*}
\item Now form $\bar{\xi^{\ast}}\in\Xi_{(N,K,b)}$ by combining blocks of size $m=2$ in $\bar{d}$ in any manner to obtain the final block design $\bar{\xi^{\ast}}\in\Xi_{(N,K,b)}$ having corresponding design matrix $\mathbf{F}$. \vspace{3mm}\par
The main effects paired comparison block design $\bar{\xi^{\ast}}$ is obtained when the profiles of the matrix $\textbf{F}$ are assigned to blocks as specified by the block indicator matrix $\mathbf{Z} = \mathbf{I}_{b}\otimes \mathbf{1}_{m}$ for which $\mathbf{F}^{\top}\mathbf{Z}=\mathbf{0}$.  \par
Now from \eqref{eq:4.47} we observe that the final design $\bar{\xi^{\ast}}\in\Xi_{(N,K,b)}$ still has information matrix of the form
\begin{equation}\label{eq:4.49}
\mathbf{M}(\bar{\xi^{\ast}})= \mathbf{F}^{\top}\mathbf{F}-(1/m)\mathbf{F}^{\top}\mathbf{Z}\mathbf{Z}^{\top}\mathbf{F} =\mathbf{F}^{\top}\mathbf{F}=(N-2)\mathbf{I}_{K}+2\mathbf{J}_{K}.
\end{equation}\par

Hence, $\bar{\xi^{\ast}}$ is $\phi_{f}$-optimal ($D$-optimal) in $\Xi_{(N,K,b)}$ \citep[e.g.,~see][Theorem~3.3]{jacroux2013optimal}.
\end{steps}

\newtheorem{exmp}{Example}
\begin{exmp}\label{eg2}
\rm{Table \ref{tab5.1} illustrates the design matrix $\textbf{F}$ with $18$ paired comparisons constructed via Method $1$ for the optimal block designs $\bar{\xi^{\ast}}$ within classes $\Xi_{(18,K,9)}$ where the main effects are orthogonal to the block effects. The matrices $\textbf{F}_1$ and $\textbf{F}_2$ are constructed by converting the attribute levels in each pair $(1,2)$ and $(2,1)$ using effects-coding whose difference yield the design matrix $\textbf{F}$ in the last column of the table.}
\end{exmp}

\begin{table}[H]
\centering
 \caption{Illustration of construction for optimal main-effects-only block designs for $K=6$ attributes with $N=18$ pairs}\label{tab5.1}
\resizebox{!}{.2\paperwidth}{
  \begin{tabular}{ccccccccc}\toprule
 Replaced attribute levels & Matrix $\textbf {F}_1$ & Matrix $\textbf {F}_2$ & $\textbf {F}=\textbf {F}_1-\textbf {F}_2$\\  
\hline
\ Level pair                  & Alternative 1       & Alternative 2      & Difference matrix\\\hline 
(1,2)\ (1,2)\ (1,2)\ (1,2)\ (1,2)\ (1,2)& \ 1 \ 1 \ 1 \ 1 \ 1 \ 1& -1  -1  -1  -1  -1  -1& \ 2 \ 2 \ 2 \ 2 \ 2 \ 2	\\
(2,1)\ (2,1)\ (2,1)\ (2,1)\ (2,1)\ (2,1)&  -1  -1  -1  -1  -1  -1&\ 1 \ 1 \ 1 \ 1 \ 1 \ 1&  -2  -2  -2  -2  -2  -2	\\
(1,2)\ (2,1)\ (1,2)\ (2,1)\ (1,2)\ (2,1)& \ 1  -1 \ 1  -1 \ 1  -1& -1 \ 1  -1 \ 1  -1 \ 1& \ 2  -2 \ 2  -2 \ 2  -2	\\
(2,1)\ (1,2)\ (2,1)\ (1,2)\ (2,1)\ (1,2)&  -1 \ 1  -1 \ 1  -1 \ 1&\ 1  -1 \ 1  -1 \ 1  -1&  -2 \ 2  -2 \ 2  -2 \ 2	\\
(1,2)\ (1,2)\ (2,1)\ (2,1)\ (1,2)\ (1,2)& \ 1 \ 1  -1  -1 \ 1 \ 1& -1  -1 \ 1 \ 1  -1  -1& \ 2 \ 2  -2  -2 \ 2 \ 2	\\
(2,1)\ (2,1)\ (1,2)\ (1,2)\ (2,1)\ (2,1)&  -1  -1 \ 1 \ 1  -1  -1&\ 1 \ 1  -1  -1 \ 1 \ 1&  -2  -2 \ 2 \ 2  -2  -2	\\
(1,2)\ (2,1)\ (2,1)\ (1,2)\ (1,2)\ (2,1)& \ 1  -1  -1 \ 1 \ 1  -1& -1 \ 1 \ 1  -1  -1 \ 1& \ 2  -2  -2 \ 2 \ 2  -2	\\
(2,1)\ (1,2)\ (1,2)\ (2,1)\ (2,1)\ (1,2)&  -1 \ 1 \ 1  -1  -1 \ 1&\ 1  -1  -1 \ 1 \ 1  -1&  -2 \ 2 \ 2  -2  -2 \ 2	\\
(1,2)\ (1,2)\ (1,2)\ (1,2)\ (2,1)\ (2,1)& \ 1 \ 1 \ 1 \ 1  -1  -1& -1  -1  -1  -1 \ 1 \ 1& \ 2 \ 2 \ 2 \ 2  -2  -2	\\
(2,1)\ (2,1)\ (2,1)\ (2,1)\ (1,2)\ (1,2)&  -1  -1  -1  -1 \ 1 \ 1&\ 1 \ 1 \ 1 \ 1  -1  -1&  -2  -2  -2  -2 \ 2 \ 2	\\
(1,2)\ (2,1)\ (1,2)\ (2,1)\ (2,1)\ (1,2)& \ 1  -1 \ 1  -1  -1 \ 1& -1 \ 1  -1 \ 1 \ 1  -1& \ 2  -2 \ 2  -2  -2 \ 2	\\
(2,1)\ (1,2)\ (2,1)\ (1,2)\ (1,2)\ (2,1)&  -1 \ 1  -1 \ 1 \ 1  -1&\ 1  -1 \ 1  -1  -1 \ 1&  -2 \ 2  -2 \ 2 \ 2  -2	\\
(1,2)\ (1,2)\ (2,1)\ (2,1)\ (2,1)\ (2,1)& \ 1 \ 1  -1  -1  -1  -1& -1  -1 \ 1 \ 1 \ 1 \ 1& \ 2 \ 2  -2  -2  -2  -2	\\
(2,1)\ (2,1)\ (1,2)\ (1,2)\ (1,2)\ (1,2)&  -1  -1 \ 1 \ 1 \ 1 \ 1&\ 1 \ 1  -1  -1  -1  -1&  -2  -2 \ 2 \ 2 \ 2 \ 2	\\
(1,2)\ (2,1)\ (2,1)\ (1,2)\ (2,1)\ (1,2)& \ 1  -1  -1 \ 1  -1 \ 1& -1 \ 1 \ 1  -1 \ 1  -1& \ 2  -2  -2 \ 2  -2 \ 2	\\
(2,1)\ (1,2)\ (1,2)\ (2,1)\ (1,2)\ (2,1)&  -1 \ 1 \ 1  -1 \ 1  -1&\ 1  -1  -1 \ 1  -1 \ 1&  -2 \ 2 \ 2  -2 \ 2  -2	\\
(1,2)\ (1,2)\ (1,2)\ (1,2)\ (1,2)\ (1,2)& \ 1 \ 1 \ 1 \ 1 \ 1 \ 1& -1  -1  -1  -1  -1  -1& \ 2 \ 2 \ 2 \ 2 \ 2 \ 2 	\\
(2,1)\ (2,1)\ (2,1)\ (2,1)\ (2,1)\ (2,1)&  -1  -1  -1  -1  -1  -1&\ 1 \ 1 \ 1 \ 1 \ 1 \ 1&  -2  -2  -2  -2  -2  -2	\\\bottomrule
\end{tabular}}
\end{table}

\section*{Method~2:}%$bk=N\equiv2$ $(\bmod \ 4)$, $b=(k/2)+2 \ odd$, $k>2$}
This method of constructing paired comparison block design $\bar{\xi^{\ast}}$ in classes $\Xi_{(N,K,b)}$ where $bm=N\equiv2$ $(\bmod \ 4)$ for odd block $b=m/2+2$ and block size $m>2$ follows directly from Method $1$. Now for the construction we use the following steps:
\begin{steps}
\item Let $d_{1}\in \Xi_{N_1,K,b_1}$ where $N_1/b_1=m$ be a blocked orthogonal main effects paired comparison design with matrix $\mathbf{F}_{d_{1}}$ of dimension $(N/2-1)\times(N/2-1)$ generated from the corresponding matrix $\textbf{F}_{\hat{d}}$ in Method $1$ having
\begin{equation*}
\mathbf{F}_{d_{1}}^{\top}\textbf{F}_{d_{1}}=N_1\mathbf{I}_K.
\end{equation*}
\item Let $d_{2}\in \Xi_{(N_2,K,b_2)}$ where $N_2/b_2=N_1/b_1=m$ be a blocked main effects paired comparison design having
\begin{equation*}
\textbf{F}_{d_{2}}^{\top}\textbf{F}_{d_{2}}=(N_2-2)\textbf{I}_{K}+2\textbf{J}_{K},
\end{equation*}
as in the corresponding Step $2$ of Method $1$.
\item Now form $\bar{\xi^{\ast}}\in \Xi_{(N,K,b)}$ where $N=N_1+N_2$ and $b=b_1+b_2$
\begin{equation*}
\textbf{F}=\begin{pmatrix}\textbf{F}_{d_{1}} \\ \textbf{F}_{d_{2}}\end{pmatrix}
\end{equation*}
where the blocks of $\bar{\xi^{\ast}}$ correspond to the blocks of $d_1$ and $d_2$. \vspace{3mm}\par
We observe that the final paired comparison block design $\bar{\xi^{\ast}}\in\Xi_{(N,K,b)}$ has information matrix $\mathbf{M}(\bar{\xi^{\ast}})$ of the same form as \eqref{eq:4.49} and is hence also $\phi_{f}$-optimal ($D$-optimal) in $\Xi_{(N,K,b)}$.
\end{steps}

Now, we give an example to illustrate the construction process of Method $2$.
\begin{exmp}\label{eg3}
\rm{Suppose an experimenter is interested in constructing an optimal block designs $\bar{\xi^{\ast}}$ in classes $\Xi_{(30,K,3)}$ for $N=30$ paired comparisons in $b=3$ blocks where the main effects is orthogonal to the block effects. Then using analogous arguments as in Example \ref{eg2} we obtain the corresponding design matrix $\textbf{F}$ in the last column of the Table \ref{tab5.2}.}
\end{exmp}
 \begin{landscape}
\begin{table}[H]
\centering
 \caption{Illustration of construction for optimal main-effects-only block designs.}\label{tab5.2}
 \resizebox{!}{.21\paperheight}{
  \begin{tabular}{ccccccccc}\toprule
 Replaced attribute levels & Matrix $\textbf {F}_1$ & Matrix $\textbf {F}_2$ & $\textbf {F}=\textbf {F}_1-\textbf {F}_2$\\  
\hline
\ Level pair                  & Alternative 1       & Alternative 2      & Difference matrix\\\hline 
(1,2)\ (1,2)\ (1,2)\ (1,2)\ (1,2)\ (1,2)& \ 1 \ 1 \ 1 \ 1 \ 1 \ 1& -1 -1 -1 -1 -1 -1& \ 2 \ 2 \ 2 \ 2 \ 2 \ 2 \\
(2,1)\ (2,1)\ (2,1)\ (2,1)\ (2,1)\ (2,1)& -1 -1 -1 -1 -1 -1&\ 1 \ 1 \ 1 \ 1 \ 1 \ 1& -2 -2 -2 -2 -2 -2 \\
(1,2)\ (2,1)\ (1,2)\ (1,2)\ (2,1)\ (2,1)& \ 1 -1 \ 1 \ 1 -1 -1& -1 \ 1 -1 -1 \ 1 \ 1& \ 2 -2 \ 2 \ 2 -2 -2 \\
(2,1)\ (1,2)\ (2,1)\ (2,1)\ (1,2)\ (1,2)& -1 \ 1 -1 -1 \ 1 \ 1&\ 1 -1 \ 1 \ 1 -1 -1& -2 \ 2 -2 -2 \ 2 \ 2 \\
(1,2)\ (1,2)\ (2,1)\ (1,2)\ (2,1)\ (1,2)& \ 1 \ 1 -1 \ 1 -1 \ 1& -1 -1 \ 1 -1 \ 1 -1& \ 2 \ 2 -2 \ 2 -2 \ 2 \\
(2,1)\ (2,1)\ (1,2)\ (2,1)\ (1,2)\ (2,1)& -1 -1 \ 1 -1 \ 1 -1&\ 1 \ 1 -1 \ 1 -1 \ 1& -2 -2 \ 2 -2 \ 2 -2 \\
(1,2)\ (2,1)\ (2,1)\ (1,2)\ (1,2)\ (2,1)& \ 1 -1 -1 \ 1 \ 1 -1& -1 \ 1 \ 1 -1 -1 \ 1& \ 2 -2 -2 \ 2 \ 2 -2 \\
(2,1)\ (1,2)\ (1,2)\ (2,1)\ (2,1)\ (1,2)& -1 \ 1 \ 1 -1 -1 \ 1&\ 1 -1 -1 \ 1 \ 1 -1& -2 \ 2 \ 2 -2 -2 \ 2 \\
(1,2)\ (1,2)\ (1,2)\ (2,1)\ (1,2)\ (2,1)& \ 1 \ 1 \ 1 -1 \ 1 -1& -1 -1 -1 \ 1 -1 \ 1& \ 2 \ 2 \ 2 -2 \ 2 -2 \\
(2,1)\ (2,1)\ (2,1)\ (1,2)\ (2,1)\ (1,2)& -1 -1 -1 \ 1 -1 \ 1&\ 1 \ 1 \ 1 -1 \ 1 -1& -2 -2 -2 \ 2 -2 \ 2 \\
(1,2)\ (2,1)\ (1,2)\ (2,1)\ (2,1)\ (1,2)& \ 1 -1 \ 1 -1 -1 \ 1& -1 \ 1 -1 \ 1 \ 1 -1& \ 2 -2 \ 2 -2 -2 \ 2 \\
(2,1)\ (1,2)\ (2,1)\ (1,2)\ (1,2)\ (2,1)& -1 \ 1 -1 \ 1 \ 1 -1&\ 1 -1 \ 1 -1 -1 \ 1& -2 \ 2 -2 \ 2 \ 2 -2 \\
(1,2)\ (1,2)\ (2,1)\ (2,1)\ (2,1)\ (2,1)& \ 1 \ 1 -1 -1 -1 -1& -1 -1 \ 1 \ 1 \ 1 \ 1& \ 2 \ 2 -2 -2 -2 -2 \\
(2,1)\ (2,1)\ (1,2)\ (1,2)\ (1,2)\ (1,2)& -1 -1 \ 1 \ 1 \ 1 \ 1&\ 1 \ 1 -1 -1 -1 -1& -2 -2 \ 2 \ 2 \ 2 \ 2 \\
(1,2)\ (2,1)\ (2,1)\ (2,1)\ (1,2)\ (1,2)& \ 1 -1 -1 -1 \ 1 \ 1& -1 \ 1 \ 1 \ 1 -1 -1& \ 2 -2 -2 -2 \ 2 \ 2 \\
(2,1)\ (1,2)\ (1,2)\ (1,2)\ (2,1)\ (2,1)& -1 \ 1 \ 1 \ 1 -1 -1&\ 1 -1 -1 -1 \ 1 \ 1& -2 \ 2 \ 2 \ 2 -2 -2 \\
(1,2)\ (1,2)\ (1,2)\ (1,2)\ (1,2)\ (1,2)& \ 1 \ 1 \ 1 \ 1 \ 1 \ 1& -1 -1 -1 -1 -1 -1& \ 2 \ 2 \ 2 \ 2 \ 2 \ 2 \\
(2,1)\ (2,1)\ (2,1)\ (2,1)\ (2,1)\ (2,1)& -1 -1 -1 -1 -1 -1&\ 1 \ 1 \ 1 \ 1 \ 1 \ 1& -2 -2 -2 -2 -2 -2 \\
(1,2)\ (2,1)\ (1,2)\ (1,2)\ (1,2)\ (2,1)& \ 1 -1 \ 1 \ 1 \ 1 -1& -1 \ 1 -1 -1 -1 \ 1& \ 2 -2 \ 2 \ 2 \ 2 -2 \\
(2,1)\ (1,2)\ (1,2)\ (2,1)\ (1,2)\ (1,2)& -1 \ 1 \ 1 -1 \ 1 \ 1&\ 1 -1 -1 \ 1 -1 -1& -2 \ 2 \ 2 -2 \ 2 \ 2 \\
(2,1)\ (2,1)\ (2,1)\ (1,2)\ (2,1)\ (1,2)& -1 -1 -1 \ 1 -1 \ 1&\ 1 \ 1 \ 1 -1 \ 1 -1& -2 -2 -2 \ 2 -2 \ 2 \\
(1,2)\ (2,1)\ (2,1)\ (2,1)\ (1,2)\ (2,1)& \ 1 -1 -1 -1 \ 1 -1& -1 \ 1 \ 1 \ 1 -1 \ 1& \ 2 -2 -2 -2 \ 2 -2 \\
(1,2)\ (1,2)\ (2,1)\ (2,1)\ (2,1)\ (1,2)& \ 1 \ 1 -1 -1 -1 \ 1& -1 -1 \ 1 \ 1 \ 1 -1& \ 2 \ 2 -2 -2 -2 \ 2 \\
(2,1)\ (1,2)\ (1,2)\ (1,2)\ (2,1)\ (2,1)& -1 \ 1 \ 1 \ 1 -1 -1&\ 1 -1 -1 -1 \ 1 \ 1& -2 \ 2 \ 2 \ 2 -2 -2 \\
(1,2)\ (1,2)\ (2,1)\ (1,2)\ (1,2)\ (1,2)& \ 1 \ 1 -1 \ 1 \ 1 \ 1& -1 -1 \ 1 -1 -1 -1& \ 2 \ 2 -2 \ 2 \ 2 \ 2 \\
(1,2)\ (2,1)\ (1,2)\ (1,2)\ (2,1)\ (1,2)& \ 1 -1 \ 1 \ 1 -1 \ 1& -1 \ 1 -1 -1 \ 1 -1& \ 2 -2 \ 2 \ 2 -2 \ 2 \\
(2,1)\ (1,2)\ (2,1)\ (1,2)\ (1,2)\ (2,1)& -1 \ 1 -1 \ 1 \ 1 -1&\ 1 -1 \ 1 -1 -1 \ 1& -2 \ 2 -2 \ 2 \ 2 -2 \\
(2,1)\ (2,1)\ (1,2)\ (2,1)\ (1,2)\ (1,2)& -1 -1 \ 1 -1 \ 1 \ 1&\ 1 \ 1 -1 \ 1 -1 -1& -2 -2 \ 2 -2 \ 2 \ 2 \\
(1,2)\ (1,2)\ (1,2)\ (2,1)\ (2,1)\ (2,1)& \ 1 \ 1 \ 1 -1 -1 -1& -1 -1 -1 \ 1 \ 1 \ 1& \ 2 \ 2 \ 2 -2 -2 -2 \\
(2,1)\ (2,1)\ (2,1)\ (2,1)\ (2,1)\ (2,1)& -1 -1 -1 -1 -1 -1&\ 1 \ 1 \ 1 \ 1 \ 1 \ 1& -2 -2 -2 -2 -2 -2 \\\bottomrule
\end{tabular}}
\end{table}
 \end{landscape}
\clearpage
\section*{Method~3:}% $n\equiv2$ $(\bmod \ 8)$, $k>2$ \rm{even}}
Additional paired comparison block design $\bar{\xi^{\ast}}$ within classes $\Xi_{(N,K,b)}$ where $N\equiv2$ $(\bmod \ 8)$ for block size $m=N/b>2$ and $m$ is even can be constructed by combining successive sets of $b=m/2$ blocks from the design matrix $\textbf{F}_{\bar{d}}$ with corresponding design $\bar{d}$ in Method $1$ to form paired comparison block design $\bar{\xi^{\ast}}$. \par
If the paired comparison block design $\bar{\xi^{\ast}}\in\Xi_{(N,K,b)}$ is constructed as in Method $1$ with corresponding design matrix $\textbf{F}$, then we observe that the final design $\bar{\xi^{\ast}}\in\Xi_{(N,K,b)}$ has information matrix $\mathbf{M}(\bar{\xi^{\ast}})$ of the form as \eqref{eq:4.49} and has eigenvalues $\lambda^{\ast}_{1}=\dots=\lambda^{\ast}_{K-1}=N-2$ and $\lambda^{\ast}_{K}=N+2K-2$. We note that the eigenvalues $\lambda^{\ast}_{1}\dots,\lambda^{\ast}_{K}$ are obtained when the profiles of the design matrix $\textbf{F}$ are assigned to blocks as specified by the block indicator matrix $\mathbf{Z} = \mathbf{I}_{b}\otimes \mathbf{1}_{m}$. Hence, $\bar{\xi^{\ast}}$ is $D$-optimal in $\Xi_{(N,K,b)}$ \citep[e.g.,~see][Theorem~3.1]{jacroux2011d} and \citet[][p.~234]{jacroux1983optimality}.

% \citep[e.g., see][p.~138]{jacroux2013d}, \citep[see also][p.~117]{jacroux2011d}. The design $\hat{d}$ is optimal by $\textbf{Theorem 2.3}$ of \citet[p.~234]{jacroux1983optimality} and $\textbf{Theorem 3.1}$ of \citet{jacroux2011d}.

\section*{Method~4:}% $n\equiv2$ $(\bmod \ 8)$, $k>2$ \rm{even}}\label{mth4}
Now we construct a non-orthogonally paired comparison block design $\bar{\xi^{\ast}}\in\Xi_{(N,K,b)}$ where $N\equiv2$ $(\bmod \ 8)$, $m=N/b>2$ even, $b<N/2$ and not all the $K$ attributes are orthogonal to the blocks. The design is constructed as follows:

%The non-orthogonally blocked design $d_4$ which is optimal according to $\textbf{Theorem 3.1}$ \citep[e.g., see][p.~140]{jacroux2013d}, \citep[see also Theorem 4.1 of][p.~118--119]{jacroux2011d} is constructed as follows:
\begin{steps}
\item Let $\textbf{L}$ consists of $K$ columns from an $(n-2)/2\times(n-2)/2$ Hadamard matrix.
\item Now let design $d_1\in\Xi_{(N-2,K,(N-2)/2)}$ have 
\begin{equation*}
\textbf{F}_{d_1}=\textbf{L}\otimes\mathbf{A}_{v}
\end{equation*}
and where blocks of size two are formed by taking successive two profiles in $\textbf{F}_{d_1}$.
\item Form design $d_2\in\Xi_{(N,K,N/2)}$ by adding a single block to $d_2$ consisiting of one profile of all $2$'s followed by another profile consisting of $K_1$ $2$'s followed by $N-K_1$ $-2$'s where $K_1>0$.
\item Form $\bar{\xi^{\ast}}\in\Xi_{(N,K,b)}$ by combining $m/2$ successive profiles of blocks of size two in $\textbf{F}_{d_2}$ to form blocks of size $m$ in $\bar{\xi^{\ast}}$ having corresponding design matrix $\mathbf{F}$. \vspace{3mm}\par
Now from \eqref{eq:4.47} we observe that the final design $\bar{\xi^{\ast}}\in\Xi_{(N,K,b)}$ has information matrix $\mathbf{M}(\bar{\xi^{\ast}})$ of the form
\begin{equation}\label{eq:4.50}
\mathbf{M}(\bar{\xi^{\ast}})= \mathbf{F}^{\top}\mathbf{F}-(1/m)\mathbf{F}^{\top}\mathbf{Z}\mathbf{Z}^{\top}\mathbf{F} =(N-2)\mathbf{I}_{K}+2\mathbf{J}_{K}-(4/m)\mathbf{J}_{K}
\end{equation}
and that $\mathbf{M}(\bar{\xi^{\ast}})$ has eigenvalues $\lambda^{\ast}_{1}=\dots=\lambda^{\ast}_{K-2}=N-2$, $\lambda^{\ast}_{K-1}=N+2K_1-2-4K_1/m$ and $\lambda^{\ast}_{K}=N+2K-2K_1-2$. \par
Hence, $\bar{\xi^{\ast}}$ is $D$-optimal in $\Xi_{(N,K,b)}$ \citep[e.g.,~see][Theorem~3.1]{jacroux2013d}.
\end{steps}

Now in the following we give an example to illustrate the construction process of non-orthogonal block designs of Method $4$.
\begin{exmp}\label{eg4}
\rm{Suppose one is interested in constructing an optimal block designs $\bar{\xi^{\ast}}$ in classes $\Xi_{(18,K,3)}$ for $N=18$ paired comparisons in $b=3$ blocks where not all the main effects are orthogonal to the block effects. Then by using similar arguments as in Example \ref{eg2} we obtain the corresponding design matrix $\textbf{F}$ in the last column of the Table \ref{tab5.3} below.}
\end{exmp}

 \begin{landscape}
\begin{table}[H]
 \caption{Illustration of construction for optimal main-effects-only block designs for $K=8$ attributes with $N=18$ pairs}\label{tab5.3}
  \begin{tabular}{ccccccccc}\toprule
 Replaced attribute levels & Matrix $\textbf {F}_1$ & Matrix $\textbf {F}_2$ & $\textbf {F}=\textbf {F}_1-\textbf {F}_2$\\  
\hline
\ Level pair                  & Alternative 1       & Alternative 2      & Difference matrix\\\hline 
(1,2)	(1,2)	(1,2)	(1,2)	(1,2)	(1,2)	(1,2)	(1,2)	& \ 1 \ 1 \ 1	 \ 1 \ 1 \ 1  \ 1 \ 1&	-1  -1	-1  -1	-1  -1   -1  -1 & \ 2 \ 2 \ 2 \ 2  \ 2  \ 2 \  2  \ 2    \\
(2,1)	(2,1)	(2,1)	(2,1)	(2,1)	(2,1)	(2,1)	(2,1)	& -1   -1  -1	  -1  -1  -1   -1  -1&\  1 \ 1 \ 1 \ 1 \ 1 \ 1	\ 1 \ 1	&  -2  -2  -2  -2   -2	 -2   -2   -2	\\
(1,2)	(2,1)	(1,2)	(1,2)	(2,1)	(2,1)	(1,2)	(2,1)	&\ 1   -1 \ 1  \   1  -1  -1  \ 1  -1&	-1 \ 1	-1  -1 \ 1 \ 1	 -1 \ 1	& \ 2  -2 \ 2 \ 2   -2	 -2  \ 2   -2	\\
(2,1)	(1,2)	(2,1)	(2,1)	(1,2)	(1,2)	(2,1)	(1,2)	& -1 \  1  -1	  -1 \ 1 \ 1   -1 \ 1&\  1  -1 \ 1 \ 1	-1  -1	\ 1  -1	&  -2 \ 2  -2  -2  \ 2	\ 2   -2  \ 2	\\
(1,2)	(1,2)	(2,1)	(1,2)	(2,1)	(1,2)	(2,1)	(2,1)	&\ 1 \  1  -1  \   1  -1 \ 1   -1  -1&	-1  -1 \ 1  -1 \ 1  -1	\ 1 \ 1	& \ 2 \ 2  -2 \ 2   -2	\ 2   -2   -2	\\
(2,1)	(2,1)	(1,2)	(2,1)	(1,2)	(2,1)	(1,2)	(1,2)	& -1   -1 \ 1	  -1 \ 1  -1 \  1 \ 1&\  1 \ 1	-1 \ 1	-1 \ 1	 -1  -1	&  -2  -2 \ 2  -2 \  2	 -2 \  2 \  2	\\
(1,2)	(2,1)	(2,1)	(1,2)	(1,2)	(2,1)	(2,1)	(1,2)	&\ 1   -1  -1 \    1 \ 1  -1   -1 \ 1&	-1 \ 1 \ 1  -1	-1 \ 1	\ 1  -1	& \ 2  -2  -2 \ 2 \  2	 -2   -2 \  2	\\
(2,1)	(1,2)	(1,2)	(2,1)	(2,1)	(1,2)	(1,2)	(2,1)	& -1 \  1 \ 1	  -1  -1 \ 1 \  1  -1&\  1  -1  -1 \ 1 \ 1  -1	 -1 \ 1	&  -2 \ 2 \ 2  -2   -2	\ 2 \  2   -2	\\
(1,2)	(1,2)	(1,2)	(2,1)	(1,2)	(2,1)	(2,1)	(2,1)	&\ 1 \  1 \ 1	  -1 \ 1  -1   -1  -1&	-1  -1	-1 \ 1	-1 \ 1	\ 1 \ 1	& \ 2 \ 2 \ 2  -2 \  2	 -2   -2   -2	\\
(2,1)	(2,1)	(2,1)	(1,2)	(2,1)	(1,2)	(1,2)	(1,2)	& -1   -1  -1 \    1  -1 \ 1 \  1 \ 1&\  1 \ 1 \ 1  -1 \ 1  -1	 -1  -1	&  -2  -2  -2 \ 2   -2	\ 2 \  2 \  2	\\
(1,2)	(2,1)	(1,2)	(2,1)	(2,1)	(1,2)	(2,1)	(1,2)	&\ 1   -1 \ 1	  -1  -1 \ 1   -1 \ 1&	-1 \ 1	-1 \ 1 \ 1  -1	\ 1  -1	& \ 2  -2 \ 2  -2   -2	\ 2   -2 \  2	\\
(2,1)	(1,2)	(2,1)	(1,2)	(1,2)	(2,1)	(1,2)	(2,1)	& -1 \  1  -1 \    1 \ 1  -1 \  1  -1&\  1  -1 \ 1  -1	-1 \ 1	 -1 \ 1	&  -2 \ 2  -2 \ 2 \  2	 -2 \  2   -2	\\
(1,2)	(1,2)	(2,1)	(2,1)	(2,1)	(2,1)	(1,2)	(1,2)	&\ 1 \  1  -1	  -1  -1  -1 \  1 \ 1&	-1  -1 \ 1 \ 1 \ 1 \ 1	 -1  -1	& \ 2 \ 2  -2  -2   -2	 -2 \  2 \  2	\\
(2,1)	(2,1)	(1,2)	(1,2)	(1,2)	(1,2)	(2,1)	(2,1)	& -1   -1 \ 1 \    1 \ 1 \ 1   -1  -1&\  1 \ 1	-1  -1	-1  -1	\ 1 \ 1	&  -2  -2 \ 2 \ 2 \  2	\ 2   -2   -2	\\
(1,2)	(2,1)	(2,1)	(2,1)	(1,2)	(1,2)	(1,2)	(2,1)	&\ 1   -1  -1	  -1 \ 1 \ 1 \  1  -1&	-1 \ 1 \ 1 \ 1	-1  -1	 -1 \ 1	& \ 2  -2  -2  -2 \  2	\ 2 \  2   -2	\\
(2,1)	(1,2)	(1,2)	(1,2)	(2,1)	(2,1)	(2,1)	(1,2)	& -1 \  1 \ 1 \    1  -1  -1   -1 \ 1&\  1  -1	-1  -1 \ 1 \ 1	\ 1  -1	&  -2 \ 2 \ 2 \ 2   -2	 -2   -2 \  2	\\
(1,2)	(1,2)	(1,2)	(1,2)	(1,2)	(1,2)	(1,2)	(1,2)	&\ 1 \  1 \ 1 \    1 \ 1 \ 1 \  1 \ 1&	-1  -1	-1  -1	-1  -1 	 -1  -1	& \ 2 \ 2 \ 2 \ 2 \  2	\ 2 \  2 \  2	\\
(1,2)	(1,2)	(2,1)	(2,1)	(2,1)	(2,1)	(2,1)	(2,1)	&\ 1 \  1  -1	  -1  -1  -1   -1  -1&	-1  -1 \ 1 \ 1 \ 1 \ 1	\ 1 \ 1	& \ 2 \ 2  -2  -2   -2	 -2   -2   -2	\\\bottomrule
\end{tabular}
\end{table}
 \end{landscape}
 
\section*{Method~5:}% $bk=N\equiv0$ $(\bmod \ 4)$, $k>1 \ odd$}
For this method of paired comparison block design $\bar{\xi^{\ast}}$ construction, let $\Xi_{(N,K;b;m)}$ be a class of designs where $bm=N\equiv0$ $(\bmod \ 4)$ and $m$ is odd. The construction is given as follows:
\begin{steps}
\item Suppose $N=bm$ and let $\textbf{H}_b$ be a $b\times b$ Hadamard matrix. Let $\textbf{H}_K$ consists of $K$ columns from $\textbf{H}_b$.
\item  Let $\boldsymbol{\iota}$ be a column vector having entries $(m+1)/2$ of $\mathbf{A}_v$ and $(m-1)/2$ of $\mathbf{A}$. Now let $\bar{\xi^{\ast}}\in\Xi_{(N,K;b;m)}$ be the paired comparison block design having design matrix 
\begin{equation*}
\mathbf{F}=\mathbf{H}_K\otimes\boldsymbol{\iota}
\end{equation*}
where the blocks of $\bar{\xi^{\ast}}$ consist of succeeding sets of $m$ profiles at $\mathbf{F}$. \vspace{3mm}\par

%The desired design $\bar{d}$ is optimal by $\textbf{Definition 2.3}$, $\textbf{Equation 2.5}$ and $\textbf{Equation 2.6}$.\citep[e.g., see][p.~364]{jacroux2015use}
Now from \eqref{eq:4.47} we observe that the final design $\bar{\xi^{\ast}}\in\Xi_{(N,K;b;m)}$ has information matrix $\mathbf{M}(\bar{\xi^{\ast}})$ of the form
\begin{equation*}
\mathbf{M}(\bar{\xi^{\ast}})= \mathbf{F}^{\top}\mathbf{F}-(1/m)\mathbf{F}^{\top}\mathbf{Z}\mathbf{Z}^{\top}\mathbf{F} =(N-b/m)\mathbf{I}_{K}
\end{equation*}
In particular, the diagonal elements of $\mathbf{M}(\bar{\xi^{\ast}})$ are obtained when the rows of the design matrix $\textbf{F}$ are assigned to blocks as specified by the block indicator matrix $\mathbf{Z} = \mathbf{I}_{b}\otimes \mathbf{1}_{m}$ and hence $\bar{\xi^{\ast}}$ is both $A$- and $D$-optimal in $\Xi_{(N,K;b;m)}$ \citep[e.g.,~see][Definition~2.3 and Inequalities~2.4--2.6]{jacroux2015use}. \par

In the following we give an example of the method of construction given in Method $5$.
\begin{exmp}\label{eg5}
\rm{Suppose an experimenter is interested in constructing a blocked design $\bar{\xi^{\ast}}$ in classes $\Xi_{(12,K;4;3)}$ having $K\leq 4$ attributes and $N=12$ paired comparisons which are to be arranged into $b=4$ blocks of size $m=3$. Then by similarly following the lines of Example \ref{eg2} the corresponding design matrix $\textbf{F}$ can be obtained from the transposed array or matrix with replaced attribute levels below:}
\end{exmp}
{\setstretch{1.25}
$\begin{array}{rrrrrrrrrrrr}
(1,2)	(2,1)(1,2)(1,2)	(2,1)	(1,2)         (1,2)	(2,1)(1,2)(1,2)(2,1)(1,2)  \\
(1,2)(2,1)	(1,2)	(2,1)(1,2)(2,1) 	(1,2)(2,1)(1,2)(2,1)(1,2)(2,1)  \\
(1,2)(2,1)	(1,2)	(1,2)	(2,1)	(1,2)          (2,1)(1,2)(2,1)(2,1)(1,2)(2,1)  \\
(1,2)	(2,1)	(1,2)(2,1)	(1,2)(2,1) 	(2,1)(1,2)	(2,1)(1,2)(2,1)(1,2) \\ 
\end{array}$
}

Here we note that for $m$ odd the main effect estimates are not orthogonal to blocks. However, for attributes $k=1,\dots,K$ each at more than two-levels $(v_k\geq3)$ the one-way layout \citep[e.g., see][Theorem~1]{grasshoff2004optimal} can be used to construct paired comparison block designs where the estimates for the main effects will be orthogonal to blocks.
\end{steps}

\section*{Method~6:}%$N\equiv0$ $(\bmod \ 8)$}%%%
For this method of constructing block main effects paired comparison designs $\bar{\xi^{\ast}}$, let $\Xi_{(N,K;b_1,b_2;m_1,m_2)}$ be a class of designs having $b_1$ blocks of size $m_1$ and $b_2$ blocks of size $m_2$ when $N\equiv0$ $(\bmod \ 8)$. We note that the different block sizes $m_1$ and $m_2$ are all even. The block design is constructed as follows:
\begin{steps}
 \item If $N=2p\equiv0$ $(\bmod \ 8)$, let $\textbf{H}_p$ be a $p\times p$ Hadamard matrix whose first row and column are all $1$'s. Select $K$ columns from $\textbf{H}_p$ and let the resulting matrix be denoted by $\hat{\textbf{H}}_K$.
\item Let $d_1$ be the design having design matrix 
\begin{equation*}
\textbf{F}_{d_1}=\hat{\textbf{H}}_K\otimes\mathbf{A}_{v}.
\end{equation*}
\item Form the desired design $\bar{\xi^{\ast}}\in\Xi_{(N,K;b_1,b_2;m_1,m_2)}$ where $b_1=b_2$ and $m_2=m_1+2$ by successively combining $m_1/2$ two complementary profiles in $d_{1}$ to form $b_1$ blocks of size $m_1$ and then successively combining $m_2/2$ two complementary profiles in $d_{1}$ to form $b_2$ blocks of size $m_2$. Let $\textbf{F}$ be the corresponding design matrix. In particular, the block main effects paired comparison design $\bar{\xi^{\ast}}$ is obtained when the rows of the design matrix $\textbf{F}$ are assigned to blocks as specified by the block indicator matrix $$\mathbf{Z}=\begin{pmatrix}\mathbf{I}_{b_{1}}\otimes \mathbf{1}_{m_{1}}  &  0 \\0& \mathbf{I}_{b_{2}}\otimes \mathbf{1}_{m_{2}} \end{pmatrix}$$
and $\mathbf{F}^{\top}\mathbf{Z}=\mathbf{0}$.  \vspace{3mm}\par
Now we observe that the final design $\bar{\xi^{\ast}}\in\Xi_{(N,K;b_1,b_2;m_1,m_2)}$ has information matrix $\mathbf{M}(\bar{\xi^{\ast}})$ of the form \eqref{eq:4.48} because estimates for the main effects are orthogonal to blocks. \par
Hence, $\bar{\xi^{\ast}}$ is both $A$- and $D$-optimal in $\Xi_{(N,K;b_1,b_2;m_1,m_2)}$ \\\citep[see][Comment~9]{jacroux2015use}.
\end{steps}

Now we give an example to illustrate the construction process of Method $6$.
\begin{exmp}\label{eg6}
\rm{Suppose an experimenter is interested in constructing an optimal block designs $\bar{\xi^{\ast}}$ in classes $\Xi_{(24,K;4,4;2,4)}$ for $N=24$ pairs in $b_1=4$ blocks of size $m_1=2$ and $b_2=4$ blocks of size $m_2=4$ where the main effects is orthogonal to the block effects. Then using analogous arguments as in Example \ref{eg2} we can obtain the corresponding design matrix $\textbf{F}$ from the array in the first column of Table \ref{tab5.4} below. In order for the array in the first column to be used in applications, we select the first component of each row of the array to form the first alternative while the second component form the second alternative as exhibited in the last column of the table.}
\end{exmp}
\begin{table}[H]
\centering
 \caption{Illustration of construction for optimal main-effects-only block designs for $K=6$ attributes and $N=24$ pairs}\label{tab5.4}
 \resizebox{!}{.25\paperheight}{
  \begin{tabular}{ccccccc}\toprule
 Array from replaced attribute levels                       & Pairs of alternatives \\  
\hline
(1,2)	(1,2)	(1,2)	(1,2)	(1,2)	(1,2)	&((1,1,1,1,1,1),(2,2,2,2,2,2))    \\
(2,1)	(2,1)	(2,1)	(2,1)	(2,1)	(2,1)	&((2,2,2,2,2,2),(1,1,1,1,1,1))	\\
(1,2)	(2,1)	(1,2)	(2,1)	(1,2)	(1,2)	&((1,2,1,2,1,1),(2,1,2,1,2,2))	\\
(2,1)	(1,2)	(2,1)	(1,2)	(2,1)	(2,1)	&((2,1,2,1,2,2),(1,2,1,2,1,1))	\\
(1,2)	(2,1)	(2,1)	(1,2)	(2,1)	(1,2)	&((1,2,2,1,2,1),(2,1,1,2,1,2))	\\
(2,1)	(1,2)	(1,2)	(2,1)	(1,2)	(2,1)	&((2,1,1,2,1,2),(1,2,2,1,2,1))	\\
(1,2)	(1,2)	(2,1)	(2,1)	(1,2)	(2,1)	&((1,1,2,2,1,2),(2,2,1,1,2,1))	\\
(2,1)	(2,1)	(1,2)	(1,2)	(2,1)	(1,2)	&((2,2,1,1,2,1),(1,1,2,2,1,2))	\\
(1,2)	(2,1)	(1,2)	(2,1)	(2,1)	(1,2)	&((1,2,1,2,2,1),(2,1,2,1,1,2))	\\
(2,1)	(1,2)	(2,1)	(1,2)	(1,2)	(2,1)	&((2,1,2,1,1,2),(1,2,1,2,2,1))	\\
(1,2)	(2,1)	(2,1)	(1,2)	(2,1)	(2,1)	&((1,2,2,1,2,2),(2,1,1,2,1,1))	\\
(2,1)	(1,2)	(1,2)	(2,1)	(1,2)	(1,2)	&((2,1,1,2,1,1),(1,2,2,1,2,2))	\\
(1,2)	(2,1)	(2,1)	(2,1)	(1,2)	(2,1)	&((1,2,2,2,1,2),(2,1,1,1,2,1))	\\
(2,1)	(1,2)	(1,2)	(1,2)	(2,1)	(1,2)	&((2,1,1,1,2,1),(1,2,2,2,1,2))	\\
(1,2)	(1,2)	(2,1)	(2,1)	(2,1)	(1,2)	&((1,1,2,2,2,1),(2,2,1,1,1,2))	\\
(2,1)	(2,1)	(1,2)	(1,2)	(1,2)	(2,1)	&((2,2,1,1,1,2),(1,1,2,2,2,1))	\\
(1,2)	(1,2)	(1,2)	(2,1)	(2,1)	(2,1)	&((1,1,1,2,2,2),(2,2,2,1,1,1))	\\
(2,1)	(2,1)	(2,1)	(1,2)	(1,2)	(1,2)	&((2,2,2,1,1,1),(1,1,1,2,2,2))	\\
(1,2)	(1,2)	(1,2)	(1,2)	(2,1)	(2,1)	&((1,1,1,1,2,2),(2,2,2,2,1,1))	\\
(2,1)	(2,1)	(2,1)	(2,1)	(1,2)	(1,2)	&((2,2,2,2,1,1),(1,1,1,1,2,2))	\\
(1,2)	(2,1)	(1,2)	(1,2)	(1,2)	(2,1)	&((1,2,1,1,1,2),(2,1,2,2,2,1))	\\
(2,1)	(1,2)	(2,1)	(2,1)	(2,1)	(1,2)	&((2,1,2,2,2,1),(1,2,1,1,1,2))	\\
(1,2)	(1,2)	(2,1)	(1,2)	(1,2)	(1,2)	&((1,1,2,1,1,1),(2,2,1,2,2,2))	\\
(2,1)	(2,1)	(1,2)	(2,1)	(2,1)	(2,1)	&((2,2,1,2,2,2),(1,1,2,1,1,1))	\\\bottomrule
\end{tabular}}
\end{table}
\newpage
\section*{Method~7:}%$N\equiv4$ $(\bmod \ 8)$}
For this method of constructing paired comparison block design $\bar{\xi^{\ast}}$, we denote $\Xi_{(N,K;b_1,b_2;m_1,m_2)}$ as a class of designs where $N\equiv4$ $(\bmod \ 8)$ having $b_1$ blocks of size $m_1$ and $b_2$ blocks of size $m_2$. We note that the different block sizes $m_1$ and $m_2$ are all even. Now we use the following steps:
\begin{steps}
 \item If $N=2p\equiv4$ $(\bmod \ 8)$, let $\textbf{H}_p$ be a $p\times p$ Hadamard matrix whose first row and column are all $1$'s. Select $K$ columns from $\textbf{H}_p$ and let the resulting matrix be denoted by $\hat{\textbf{H}}_K$.
\item Let $\tilde{\textbf{H}}_K$ be obtained from $\hat{\textbf{H}}_K$ by adding one row of all $1$'s and one row of $[\frac{K}{2}]$ $1$'s and $K-[\frac{K}{2}]$ $-1$'s to $\hat{\textbf{H}}_K$ where $[x]$ denotes the integer part of the decimal expansion for $x\geq0$ and let $\tilde{d}$ be the design having design matrix 
\begin{equation*}
\textbf{F}_{\tilde{d}}=\tilde{\textbf{H}}_K\otimes\mathbf{A}_{v}.
\end{equation*}
\item Form the final design $\bar{\xi^{\ast}}\in\Xi_{(N,K;b_1,b_2;m_1,m_2)}$ where $b_1=b_2$ and $m_2=m_1+2$ having design matrix $\textbf{F}$ by successively combining $m_1/2$ two complementary profiles in $\tilde{d}$ to form $b_1$ blocks of size $m_1$ and then successively combining $m_2/2$ two complementary profiles in $\tilde{d}$ to form $b_2$ blocks of size $m_2$.   \vspace{3mm}\par

Now we observe that the design $\bar{\xi^{\ast}}\in\Xi_{(N,K;b_1,b_2;m_1,m_2)}$ has information matrix $\mathbf{M}(\bar{\xi^{\ast}})$ of the form
\begin{equation}\label{eq:4.51}
\mathbf{M}(\bar{\xi^{\ast}})= \mathbf{F}^{\top}\mathbf{F}-(1/m)\mathbf{F}^{\top}\mathbf{Z}\mathbf{Z}^{\top}\mathbf{F} =(N-4)\mathbf{I}_{K}+4\mathbf{J}_{K}
\end{equation}
and that when $K$ is odd $\mathbf{M}(\bar{\xi^{\ast}})$ has eigenvalues $\lambda^{\ast}_{1}=\dots=\lambda^{\ast}_{K-2}=N-4$, $\lambda^{\ast}_{K-1}=N+4K_1-2-4$ and $\lambda^{\ast}_{K}=N+4K-4K_1-4$. Moreover, when $K$ is even $\mathbf{M}(\bar{\xi^{\ast}})$ has eigenvalues $\lambda^{\ast}_{1}=\dots=\lambda^{\ast}_{K-2}=N-4$ and $\lambda^{\ast}_{K-1}=\lambda^{\ast}_{K}=N+4K_1-4$. In particular, the eigenvalues $\lambda^{\ast}_{1}\dots,\lambda^{\ast}_{K}$ are obtained when the rows of the design matrix $\textbf{F}$ are assigned to blocks as specified by the block indicator matrix $$\mathbf{Z}=\begin{pmatrix}\mathbf{I}_{b_{1}}\otimes \mathbf{1}_{m_{1}}  &  0 \\0& \mathbf{I}_{b_{2}}\otimes \mathbf{1}_{m_{2}} \end{pmatrix}$$ and $\mathbf{F}^{\top}\mathbf{Z}=\mathbf{0}$. Hence, $\bar{\xi^{\ast}}$ is both $A$- and $D$-optimal in $\Xi_{(N,K;b_1,b_2;m_1,m_2)}$ \citep[see][Comment~9]{jacroux2015use}.
\end{steps}

\section*{Method~8:}% $bk=n\equiv3$ $(\bmod \ 4)$, $k\equiv1$ $(\bmod \ 8), b=3$}
This method of constructing paired comparison block design $\bar{\xi^{\ast}}$ within classes $\Xi_{(N,K,b)}$ where $bm=N\equiv3$ $(\bmod \ 4)$, $m\equiv1$ $(\bmod \ 8), b=3$ is given in the following steps:
\begin{steps}
\item Select $(n-3)/2\times(n-3)/2$ Hadamard matrix $\textbf{H}$.
\item Let $\mathbf{L}1=\textbf{H}\otimes\mathbf{A}_{v}$ and select $K$ columns from $\mathbf{L}1$ letting the resulting matrix be denoted by $\mathbf{L}2$.
\item Form $3$ blocks of size $m-1$ with each block consisting of $(m-1)/2$ complementary two profiles from $\mathbf{L}2$.
 \item Obtain the desired design $\bar{\xi^{\ast}}$ having design matrix $\mathbf{F}$ by adding one profile of all values corresponding to the prespecified $\mathbf{A}$ to each block formed in Step $3$. \vspace{3mm}\par
 The final design $\bar{\xi^{\ast}}\in\Xi_{(N,K,b)}$ obtained in the construction process has information matrix 
\begin{equation}\label{eq:4.52}
\mathbf{M}(\bar{\xi^{\ast}})= \mathbf{F}^{\top}\mathbf{F}-(1/m)\mathbf{F}^{\top}\mathbf{Z}\mathbf{Z}^{\top}\mathbf{F} =(N-3)\mathbf{I}_{K}+3\mathbf{J}_{K}-(b/m)\mathbf{J}_{K}
\end{equation}
and that $\mathbf{M}(\bar{\xi^{\ast}})$ has eigenvalues $\lambda^{\ast}_{1}=\dots=\lambda^{\ast}_{K-1}=N-3$ and $\lambda^{\ast}_{K}=(N-3)+3K-Kb/m$. In particular, the eigenvalues $\lambda^{\ast}_{1}\dots,\lambda^{\ast}_{K}$ are obtained when the rows of the design matrix $\textbf{F}$ are assigned to blocks as specified by the block indicator matrix $\mathbf{Z} = \mathbf{I}_{b}\otimes \mathbf{1}_{m}$ and hence $\bar{\xi^{\ast}}$ is $E$-optimal in $\Xi_{(N,K,b)}$ \citep[e.g.,~see][Theorem~4.3]{jacroux2014optimality}.
 %The construction of the design $d^*\in\bar{D}(n,m,b)$, which is optimal by \textbf{Theorem 4.3} is as follows \citep[e.g., see][p.~146]{jacroux2014optimality}
\end{steps}

\section*{Method~9:}% \rm{Construction of designs with blocks of both even and odd sizes}.}
For this method of paired comparison block design construction we denote the class of designs as $\Xi_{(N,K;b_1,\dots,b_t,m_1,\dots,m_t)}$ for $i=1,\dots,t$ having $b_1$ blocks of size $m_1$, $b_2$ blocks of size $m_2$, in that order up to $b_t$ blocks of size $m_t$. Let $\bar{\xi^{\ast}}$ denote the paired comparison block design within $\Xi_{(N,K;b_1,\dots,b_t,m_1,\dots,m_t)}$. Let $\textbf{p}_i$ denotes a $m_i\times 1$ vector having $m_i/2$ positive values of $\mathbf{A}$ and $m_i/2$ negative values of $\mathbf{A}$ if $m_i$ is even and having $(m_i+1)/2$ positive values of $\mathbf{A}$ and $(m_i-1)/2$ negative values of $\mathbf{A}$ if $m_i$ is odd. Here it is assumed that for $\bar{\xi^{\ast}}\in\Xi_{(N,K;b_1,\dots,b_t,m_1,\dots,m_t)}$ the $N$ pairs are arranged so that the first $b_1m_1=N_1$ profiles form the design matrix $\textbf{F}_{d_1}$, the next $b_2m_2=N_2$ profiles form the design matrix $\textbf{F}_{d_2}$, etc. The design is constructed as follows:
\begin{steps}
\item Let $\textbf{H}_{b_i}$ be a $b_i\times b_i$ Hadamard matrix and let $\textbf{L}_{b_i}$ consist of $K$ columns of $\textbf{H}_{b_i}$, $i=1,\dots,t$. Now let $d_i\in\Xi_{(N,K;b_1,\dots,b_t,m_1,\dots,m_t)}$
\begin{equation*}
\textbf{F}_{d_i}=\textbf{L}_{b_i} \otimes\textbf{p}_{i},
\end{equation*}
where $\textbf{F}_{d_i}$ are sub-design matrix.
% Here it is assumed that for $\bar{\xi^{\ast}}\in\Xi_{N,K;b_1,\dots,b_t,m_1,\dots,m_t}$ the $N$ profiles are arranged so that the first $b_1m_1=N_1$ profiles form the design matrix $\textbf{F}_{d_1}$
\item Now let the design matrix $\textbf{F}^{\top}=(\textbf{F}^{\top}_{d_1},\textbf{F}^{\top}_{d_2},\dots,\textbf{F}^{\top}_{d_t})$. Then\\ $\bar{\xi^{\ast}}\in\Xi_{(N,K;b_1,\dots,b_t,m_1,\dots,m_t)}$ is the desired design. We note that the design $\bar{\xi^{\ast}}$ is orthogonally blocked when the rows of the design matrix $\textbf{F}$ are assigned to blocks as specified by the block indicator matrix $\mathbf{Z} = \mathbf{I}_{b_i}\otimes \mathbf{1}_{m_i}$, $i=1,\dots,t$.
\vspace{3mm}\par

We obseve that the final design $\bar{\xi^{\ast}}\in\Xi_{(N,K;b_1,\dots,b_t,m_1,\dots,m_t)}$ has information matrix 
\begin{align}\label{eq:4.53}
\mathbf{M}(\bar{\xi^{\ast}})&= \mathbf{F}^{\top}\mathbf{F}-\mathbf{F}^{\top}\mathbf{Z}(\mathbf{Z}^{\top}\mathbf{Z})^{-1}\mathbf{Z}^{\top}\mathbf{F}  \nonumber \\
&=\sum_{i=1}^{l}(N_i-b_i/m_i)\mathbf{I}_{K}+\sum_{i=l+1}^{t}N_i\mathbf{I}_{K}\nonumber \\
&=\sum_{i=1}^{t}\mathbf{M}(d_i),
\end{align}
where $\mathbf{M}(d_i)$ is the information matrix for main effects under sub-design matrix $\textbf{F}_{d_i}$ of $\textbf{F}$ having $b_i$ blocks of size $m_i$, $i=1,\dots,t$ and that $\mathbf{M}(\bar{\xi^{\ast}})$ has maximal trace among all block designs in $\Xi_{(N,K;b_1,\dots,b_t,m_1,\dots,m_t)}$.
\par
Hence, $\bar{\xi^{\ast}}$ is type $I$-optimal in $\Xi_{(N,K;b_1,\dots,b_t,m_1,\dots,m_t)}$ \\\citep[e.g.,~see][Theorem~3.1~and~Equation~(2.5)]{jacroux2015type}.
%The construction of the design $d^*\in\bar{D}(n,m;b_1,\dots,b_t,k_1,\dots,k_t)$, which is optimal according to $\textbf{Theorem 3.1}$ \citep[see][p.~41]{jacroux2015type} is
\end{steps}

\section*{Method~10:}% $n=b_1k_1+i\equiv i$ $(\bmod \ 4), i=1,2$ \rm{or} 3, $b_1\equiv0$ $(\bmod \ 4)$.}
For this method of constructing main effects paired comparison block design $\bar{\xi^{\ast}}$, we denote the class of designs as $\Xi_{(b_1m_1+i,K;im_1+i,b_1-2i-im_1;m_1+1,m_1)}$  having $N=b_1m_1+i$ pairs, $im_1+i$ blocks of size $m_1+1$ and $b_1-2i-im_1$ blocks of size $m_1$. The block design is constructed as follows:
\begin{steps}
\item Denote by $\textbf{H}_{b1}$ a $b_1\times b_1$ Hadamard matrix whose first row and column are all $1$'s and let $\textbf{L}_{b_1}$ consist of $K$ columns from $\textbf{H}_{b_1}$ such that $\textbf{L}_{b_1}$ has its first $i$ rows consisting of all $+1$'s, $i=1,2$ or $3$. Now let
\begin{equation*}
\textbf{F}_{d_1}=\textbf{L}_{b_1} \otimes\textbf{p}_{1},
\end{equation*}
where $\textbf{p}_{1}$ is a $m_1\times 1$ vector having entries of $\mathbf{A}_v$ with $m_1$ even. Observe that design $d_1\in \Xi_{(b_1m_1,K;b_1;m_1)}$ with succeeding sets of $m_1$ profiles in $\textbf{F}_{d_1}$ forming blocks. Also observe that blocks $1,\dots,i$ of $d_1$ consist of profiles of $\textbf{F}_{d_1}$ which are either all $2$'s or $-2$'s.
\item Take each of the first $im_1$ profiles of $\textbf{F}_{d_1}$ consisting of all $2$'s or all $-2$'s and add them to one of the succeeding $im_1$ blocks of $d_1$ to form design $d_2\in\Xi_{(b_1m_1,K;im_1,b_1-i-im_1;m_1+1,m_1)}$ having corresponding design matrix $\textbf{F}_{d_2}$.
\item For the appropraite value of $i=1,2$ or $3$, take $i$ additional profiles of all $2$'s and add each such profile to a block of $d_2$ of size $m_1$ to form $\bar{\xi^{\ast}}$ having corresponding design matrix $\textbf{F}$. In particular, the design $\bar{\xi^{\ast}}\in\Xi_{(b_1k_1+i,K;im_1+i,b_1-2i-im_1;m_1+1,m_1)}$ is blocked when the rows of the design matrix $\textbf{F}$ are assigned to blocks as specified by the block indicator matrix 
  $$\mathbf{Z}=\begin{pmatrix}\mathbf{I}_{im_1+i}\otimes \mathbf{1}_{m_1+1}  &  0 \\0& \mathbf{I}_{b_1-2i-im_1}\otimes \mathbf{1}_{m_{1}} \end{pmatrix}$$ and $\mathbf{F}^{\top}\mathbf{Z}=\mathbf{0}$.    \vspace{3mm}\par
Now we obseve that the final design $\bar{\xi^{\ast}}\in\Xi_{(b_1k_1+i,K;im_1+i,b_1-2i-im_1;m_1+1,m_1)}$ has information matrix 
\begin{align}\label{eq:4.54}
\mathbf{M}(\bar{\xi^{\ast}})&= \mathbf{F}^{\top}\mathbf{F}-\mathbf{F}^{\top}\mathbf{Z}(\mathbf{Z}^{\top}\mathbf{Z})^{-1}\mathbf{Z}^{\top}\mathbf{F} =\mathbf{M}(d_1)+\mathbf{M}(d_2) \nonumber \\
&=((i+m_1+i)(m_1+1)-i)\mathbf{I}_{K}+i\mathbf{J}_{K}  \nonumber \\
&\qquad+(b_1-2i-im_1)m_1\mathbf{I}_{K}-((im_1+i)/(m_i+1))\mathbf{J}_{K}\nonumber \\
&=b_1m_1\mathbf{I}_{K},
\end{align}
where $\mathbf{M}(d_1)$ is the information matrix for main effects under subdesign matrix $\textbf{F}_{d_1}$ of $\textbf{F}$ having $im_1+i$ blocks of size $m_1+1$, $\mathbf{M}(d_2)$ is the information matrix for main effects under subdesign matrix $\textbf{F}_{d_2}$ of $\textbf{F}$ having $b_1-2i-im_1$ blocks of size $m_1$ and that $\mathbf{M}(\bar{\xi^{\ast}})$ has maximal trace among all block designs in $\Xi_{(b_1k_1+i,K;im_1+i,b_1-2i-im_1;m_1+1,m_1)}$. \par
Hence $\bar{\xi^{\ast}}$ is type $I$-optimal in $\Xi_{(b_1k_1+i,K;im_1+i,b_1-2i-im_1;m_1+1,m_1)}$ \\\citep[e.g.,~see][Theorem~3.3]{jacroux2015type}.
%which is optimal according to $\textbf{Theorem 3.3}$ \citep[e.g., see][p.~41]{jacroux2015type}
\end{steps}

 \newpage
\section*{Method~11:}% $n\equiv2$ $(\bmod \ 4)$ \rm{where} $n=8q+2$, \rm{for some integer} $q\geq1$.}
For this method of main effects paired comparison block designs $\bar{\xi^{\ast}}$ construction, we define the $b$ blocks as having block sizes $m_1,m_2,\dots,m_b$ (even) and denote the corresponding class of designs as $\Xi_{(N,K,b,m_1,m_2,\dots,m_b)}$ where $N=8q+2\equiv2$ $(\bmod \ 4)$ for some integer $q\geq1$. The construction is given in the underlying steps:
\begin{steps}
\item Let the Hadamard matrix $\textbf{H}_{\frac{n}{2}-1}$ exist and let
\begin{equation*}
\textbf{L}_1=\begin{pmatrix}\textbf{H}_{\frac{n}{2}-1} \\ \textbf{1}^{\prime}_{\frac{n}{2}-1}\end{pmatrix}\otimes\mathbf{A}_v
\end{equation*}
where $\textbf{1}^{\prime}_{\frac{n}{2}-1}$ is a vector of order $1\times\frac{n}{2}-1$ with all elements one.
\item Now let $\tilde{d}$ be the design having design matrix $\textbf{F}_{\tilde{d}}$ formed from $\textbf{L}_1$ by selecting any $K$ $(1\leq K \leq \frac{n}{2}-1)$ columns.
\item Obtain the block design $\bar{\xi^{\ast}}\in\Xi_{(N,K,b,m_1,m_2,\dots,m_b)}$ having design matrix $\textbf{F}$ formed from $\textbf{F}_{\tilde{d}}$ where sets of successive $m_i$ profiles correspond to the $i^{th}$ block of $\bar{\xi^{\ast}}$, $i=1,2,\dots,b$. \vspace{3mm}\par
We observe that $\bar{\xi^{\ast}}$ has information matrix $\mathbf{M}(\bar{\xi^{\ast}})$ of the same form as \eqref{eq:4.49} and is hence optimal with respect to any generalized criterion of type $1$ ($E$-criterion) \citep[e.g., see][Theorem~2]{saharay2016optimality}.
%of the design $d^*$, which is optimal in $\bar{D}(n,m,b,k_1,k_2,\dots,k_b)$ by $\textbf{Theorem 2}$ \citep[e.g., see][p.~562]{saharay2016optimality}
\end{steps}

We give an example to illustrate the construction process of Method $11$.
\begin{exmp}\label{eg11}
\rm{Table \ref{mth11} shows an orthogonally blocked main effects design $\bar{\xi^{\ast}}$ in classes $\Xi_{(26,K,6;4,4,4,4,4,6)}$ for $N=26$ paired comparisons partitioned into $b=6$ blocks of different even block sizes $m_1,\dots,m_{b-1} =4$ and $m_b=6$. The corresponding design matrix $\textbf{F}$ in the last column of Table \ref{mth11} can be obtained using similar arguments as before in Example \ref{eg2}.}
\end{exmp}

\begin{table}[H]
\centering
 \caption{Illustration of construction for optimal main-effects-only block designs for $K=6$ attributes with $N=26$ pairs}\label{mth11}
 \resizebox{!}{.22\paperheight}{
  \begin{tabular}{ccccccccc}\toprule
 Replaced attribute levels & Matrix $\textbf {F}_1$ & Matrix $\textbf {F}_2$ & $\textbf {F}=\textbf {F}_1-\textbf {F}_2$\\  
\hline
\ Level pair                  & Alternative 1       & Alternative 2      & Difference matrix\\\hline 
(1,2)(1,2)(1,2)(1,2)(1,2)(1,2)&\ 1 \ 1 \ 1 \ 1 \ 1 \ 1&	-1	-1	-1	-1	-1	-1&\ 2 \ 2 \ 2 \ 2 \ 2 \ 2\\
(2,1)(2,1)(2,1)(2,1)(2,1)(2,1)& -1  -1  -1  -1  -1  -1&\ 1 \ 1 \ 1 \ 1 \ 1 \ 1& -2  -2  -2  -2  -2  -2\\
(1,2)(2,1)(1,2)(2,1)(1,2)(1,2)&\ 1  -1 \ 1  -1 \ 1 \ 1&	-1 \ 1	-1 \ 1	-1	-1&\ 2  -2 \ 2  -2 \ 2 \ 2\\
(2,1)(1,2)(2,1)(1,2)(2,1)(2,1)& -1 \ 1  -1 \ 1  -1  -1&\ 1	-1 \ 1	-1 \ 1 \ 1& -2 \ 2  -2 \ 2  -2  -2\\
(1,2)(2,1)(2,1)(1,2)(2,1)(1,2)&\ 1  -1  -1 \ 1  -1 \ 1&	-1 \ 1 \ 1	-1 \ 1	-1&\ 2  -2  -2 \ 2  -2 \ 2\\
(2,1)(1,2)(1,2)(2,1)(1,2)(2,1)& -1 \ 1 \ 1  -1 \ 1  -1&\ 1	-1	-1 \ 1	-1 \ 1& -2 \ 2 \ 2  -2 \ 2  -2\\
(1,2)(1,2)(2,1)(2,1)(1,2)(2,1)&\ 1 \ 1  -1  -1 \ 1  -1&	-1	-1 \ 1 \ 1	-1 \ 1&\ 2 \ 2  -2  -2 \ 2  -2\\
(2,1)(2,1)(1,2)(1,2)(2,1)(1,2)& -1  -1 \ 1 \ 1  -1 \ 1&\ 1 \ 1	-1	-1 \ 1	-1& -2  -2 \ 2 \ 2  -2 \ 2\\
(1,2)(2,1)(1,2)(2,1)(2,1)(1,2)&\ 1  -1 \ 1  -1  -1 \ 1&	-1 \ 1	-1 \ 1 \ 1	-1&\ 2  -2 \ 2  -2  -2 \ 2\\
(2,1)(1,2)(2,1)(1,2)(1,2)(2,1)& -1 \ 1  -1 \ 1 \ 1  -1&\ 1	-1 \ 1	-1	-1 \ 1& -2 \ 2  -2 \ 2 \ 2  -2\\
(1,2)(2,1)(2,1)(1,2)(2,1)(2,1)&\ 1  -1  -1 \ 1  -1  -1&	-1 \ 1 \ 1	-1 \ 1 \ 1&\ 2  -2  -2 \ 2  -2  -2\\
(2,1)(1,2)(1,2)(2,1)(1,2)(1,2)& -1 \ 1 \ 1  -1 \ 1 \ 1&\ 1	-1	-1 \ 1	-1	-1& -2 \ 2 \ 2  -2 \ 2 \ 2\\
(1,2)(2,1)(2,1)(2,1)(1,2)(2,1)&\ 1  -1  -1  -1 \ 1  -1&	-1 \ 1 \ 1 \ 1	-1 \ 1&\ 2  -2  -2  -2 \ 2  -2\\
(2,1)(1,2)(1,2)(1,2)(2,1)(1,2)& -1 \ 1 \ 1 \ 1  -1 \ 1&\ 1	-1	-1	-1 \ 1	-1& -2 \ 2 \ 2 \ 2  -2 \ 2\\
(1,2)(1,2)(2,1)(2,1)(2,1)(1,2)&\ 1 \ 1  -1  -1  -1 \ 1&	-1	-1 \ 1 \ 1 \ 1	-1&\ 2 \ 2  -2  -2  -2 \ 2\\
(2,1)(2,1)(1,2)(1,2)(1,2)(2,1)& -1  -1 \ 1 \ 1 \ 1  -1&\ 1 \ 1	-1	-1	-1 \ 1& -2  -2 \ 2 \ 2 \ 2  -2\\
(1,2)(1,2)(1,2)(2,1)(2,1)(2,1)&\ 1 \ 1 \ 1  -1  -1  -1&	-1	-1	-1 \ 1 \ 1 \ 1&\ 2 \ 2 \ 2  -2  -2  -2\\
(2,1)(2,1)(2,1)(1,2)(1,2)(1,2)& -1  -1  -1 \ 1 \ 1 \ 1&\ 1 \ 1 \ 1	-1	-1	-1& -2  -2  -2 \ 2 \ 2 \ 2\\
(1,2)(1,2)(1,2)(1,2)(2,1)(2,1)&\ 1 \ 1 \ 1 \ 1  -1  -1&	-1	-1	-1	-1 \ 1 \ 1&\ 2 \ 2 \ 2 \ 2  -2  -2\\
(2,1)(2,1)(2,1)(2,1)(1,2)(1,2)& -1  -1  -1  -1 \ 1 \ 1&\ 1 \ 1 \ 1 \ 1	-1	-1& -2  -2  -2  -2 \ 2 \ 2\\
(1,2)(2,1)(1,2)(1,2)(1,2)(2,1)&\ 1  -1 \ 1 \ 1 \ 1  -1&	-1 \ 1	-1	-1	-1 \ 1&\ 2  -2 \ 2 \ 2 \ 2  -2\\
(2,1)(1,2)(2,1)(2,1)(2,1)(1,2)& -1 \ 1  -1  -1  -1 \ 1&\ 1	-1 \ 1 \ 1 \ 1	-1& -2 \ 2  -2  -2  -2 \ 2\\
(1,2)(1,2)(2,1)(1,2)(1,2)(1,2)&\ 1 \ 1  -1 \ 1 \ 1 \ 1&	-1	-1 \ 1	-1	-1	-1&\ 2 \ 2  -2 \ 2 \ 2 \ 2\\
(2,1)(2,1)(1,2)(2,1)(2,1)(2,1)& -1  -1 \ 1  -1  -1  -1&\ 1 \ 1	-1 \ 1 \ 1 \ 1& -2  -2 \ 2  -2  -2  -2\\
(1,2)(1,2)(1,2)(1,2)(1,2)(1,2)&\ 1 \ 1 \ 1 \ 1 \ 1 \ 1&	-1	-1	-1	-1	-1	-1&\ 2 \ 2 \ 2 \ 2 \ 2 \ 2\\
(2,1)(2,1)(2,1)(2,1)(2,1)(2,1)& -1  -1  -1  -1  -1  -1&\ 1 \ 1 \ 1 \ 1 \ 1 \ 1& -2  -2  -2  -2  -2  -2\\\bottomrule
\end{tabular}}
\end{table}

\section*{Method~12:}%$n\equiv2$ $(\bmod \ 4)$ \rm{where} $n=8q+6$, \rm{for some integer} $q\geq1$.}
This method of constructing main effects paired comparison block designs $\bar{\xi^{\ast}}$ is analogous to Method $11$. We similarly denote the $b$ blocks as having block sizes $m_1,m_2,\dots,m_b$, $m_{i}$'s even and denote the corresponding class of designs as $\Xi_{(N,K,b,m_1,m_2,\dots,m_b)}$ where $N=8q+6\equiv2$ $(\bmod \ 4)$ for some integer $q\geq1$. Now we use the following steps:
\begin{steps}
\item Let the Hadamard matrix $\textbf{H}_{\frac{n}{2}+1}$ exist and can be formulated as follows:
\begin{equation*}
\textbf{H}^*_{\frac{n}{2}+1} =\begin{pmatrix}\textbf{H}_{\frac{n}{2}+1} \\ \textbf{1}^{\prime}_{\frac{n}{2}+1}\end{pmatrix},
\end{equation*}
where $\textbf{1}^{\prime}_{\frac{n}{2}+1}$ is a vector of order $1\times\frac{n}{2}+1$ with all elements one.
\item Now let 
\begin{equation*}
\textbf{L}_1=\textbf{H}^*_{\frac{n}{2}+1}\otimes\mathbf{A}_v,
\end{equation*}
by selecting any $K$ $(1\leq K \leq \frac{n}{2}+1)$ columns.
\item Obtain $\bar{\xi^{\ast}}\in\Xi_{(N,K,b,m_1,m_2,\dots,m_b)}$ having design matrix $\textbf{F}$ formed from $\textbf{L}_1$ where sets of successive $m_i$, $i,=1,2,\dots,b$ profiles correspond to the $i^{th}$ block of $\bar{\xi^{\ast}}$.  \vspace{3mm}\par
 The final design $\bar{\xi^{\ast}}\in\Xi_{(N,K,b,m_1,m_2,\dots,m_b)}$ obtained in the construction process has information matrix 
\begin{equation}\label{eq:4.55}
\mathbf{M}(\bar{\xi^{\ast}})= \mathbf{F}^{\top}\mathbf{F}-(1/m_i)\mathbf{F}^{\top}\mathbf{Z}\mathbf{Z}^{\top}\mathbf{F} =(N+2)\mathbf{I}_{K}-2\mathbf{J}_{K}.
\end{equation}
 In particular, the designs $\bar{\xi^{\ast}}$ are obtained when the rows of the design matrix $\textbf{F}$ are assigned to blocks as specified by the block indicator matrix $\mathbf{Z} = \mathbf{I}_{b}\otimes \mathbf{1}_{m_1,m_2,\dots,m_b}$ and that $\bar{\xi^{\ast}}$ is optimal in $\Xi_{(N,K,b,m_1,m_2,\dots,m_b)}$ with respect to any generalized criterion of type $2$ \citep[e.g.,~see][Theorem~4]{saharay2016optimality}.
\end{steps}

\section*{Method~13:}% $n\equiv2$ $(\bmod \ 4)$ \rm{having different even block sizes}.}
Analogous to Method 11. Here this method of constructing main effects paired comparison block design $\bar{\xi^{\ast}}$ in classes $\Xi_{(N,K,b,m_1,m_2,\dots,m_b)}$ where $N\equiv2$ $(\bmod \ 4)$ and $K\geq3$ is given in the following steps:
\begin{steps}
\item Let $\textbf{B}_1$ be the $(n-2)\times(n-2)$ matrix given by
\begin{equation*}
\textbf{B}_1=\textbf{H}_{\frac{n}{2}-1}\otimes\mathbf{A}_v,
\end{equation*}
where the matrix $\textbf{H}_{\frac{n}{2}-1}$ is defined in Method $11$.
\item Let
\begin{equation*}
\textbf{B}_2=\begin{pmatrix} \textbf{B}_1\\ \textbf{J}_{2\times \frac{n}{2}-1}\end{pmatrix},
\end{equation*}
where $\textbf{J}_{2\times \frac{n}{2}-1}$ is a ${2\times \frac{n}{2}-1}$ matrix of all elements corresponding to the values in $\mathbf{A}$.
\item Now obtain $\bar{\xi^{\ast}}\in\Xi_{(N,K,b,m_1,m_2,\dots,m_b)}$ having matrix $\textbf{F}$ formed from $\textbf{B}_2$ by selecting any $K$ $(3\leq K \leq \frac{n}{2}-1)$ columns, where sets of successive $m_i$ profiles correspond to the $i^{th}$ block of $\bar{\xi^{\ast}}$, $i=1,2,\dots,b$.  \vspace{3mm}\par
We observe that $\bar{\xi^{\ast}}$ has information matrix $\mathbf{M}(\bar{\xi^{\ast}})$ of the same form as \eqref{eq:4.49}. Then $\bar{\xi^{\ast}}$ is $E$-optimal in $\Xi_{(N,K,b,m_1,m_2,\dots,m_b)}$ \citep[e.g., see][Theorem~3]{saharay2016optimality}.
%$\textbf{Theorem 3}$ \citep[e.g., see][p.~562~-~563]{saharay2016optimality} is as follows:
\end{steps}

\section*{Method~14:}%$n=k^2\equiv1$ $(\bmod \ 4)$, $k\geq 3$.}
For this method of construction, let the desired design $\bar{\xi^{\ast}}$ in classes $\Xi_{(N,K,b)}$ be denoted as $\bar{\xi^{\ast}}\in \Xi_{(N,K,b)}$ where the block size $m\geq 3$ is odd, $N=m^2\equiv1$ $(\bmod \ 4)$ and $b=m$. Now the construction is presented in the following steps:
\begin{steps}
\item Let $\textbf{H}_{\frac{n-1}{2}}$ be a Hadamard matrix of dimension $\frac{n-1}{2}\times\frac{n-1}{2}$ and contain $(b-1)/2$ rows that are identical. Now let $\textbf{H}_{\frac{n-1}{2}}$ be partitioned so that 
 
\begin{equation*}
\textbf{H}_{\frac{n-1}{2}}=\begin{pmatrix} \textbf{H}_{1}\\ \textbf{H}_{2}\end{pmatrix}
\end{equation*}
where $\textbf{H}_{1}=\textbf{J}_{(b-1)/2}$ and $\textbf{J}_{pq}$ denotes a $p\times q$ matrix of ones.
\item Let the design $\tilde{d}\in \Xi_{(N-1,K)}$ have matrix
\begin{equation*}
\textbf{F}_{\tilde{d}}= \begin{pmatrix} \textbf{H}_{1}\\ \textbf{H}_{2}\end{pmatrix}\otimes\mathbf{A}_v.
\end{equation*}
\item Using the profiles of $\textbf{F}_{\tilde{d}}$ corresponding to $\textbf{H}_{2}\otimes\mathbf{A}_v$, form $\textbf{F}_{\hat{d}}$ having $b$ blocks of size $m-1$ where the blocks of $\textbf{F}_{\hat{d}}$ are obtained by combining successive sets of $m-1$ profiles in $\textbf{H}_{2}\otimes\mathbf{A}_v$.
\item Now form $\bar{\xi^{\ast}}$ having design matrix $\textbf{F}$ with $b$ blocks of size $m$ by taking the $b-1$ profiles of all $2$'s or $-2$'s in $\textbf{H}_{1}\otimes\mathbf{A}_v$ and add one of each such profile to each of the first $b-1$ blocks in $\textbf{F}_{\hat{d}}$ and form the last block of $\textbf{F}$ by adding a profile of all $2$'s to the last block of $\textbf{F}_{\hat{d}}$.  \vspace{3mm}\par
We note that the block design $\bar{\xi^{\ast}}$ is obtained when the profiles of the design matrix $\textbf{F}$ are assigned to blocks as specified by the block indicator matrix $\mathbf{Z} = \mathbf{I}_{b}\otimes \mathbf{1}_{m}$. As a consequence, the final design $\bar{\xi^{\ast}}\in\Xi_{(N,K,b)}$ has information matrix 
\begin{align}\label{eq:4.56}
\mathbf{M}(\bar{\xi^{\ast}})&= \mathbf{F}^{\top}\mathbf{F}-(1/m)\mathbf{F}^{\top}\mathbf{Z}\mathbf{Z}^{\top}\mathbf{F}    \\
&=(N-1)\mathbf{I}_{K}+\mathbf{J}_{K}-(b/m)\mathbf{J}_{K}   \nonumber\\
&=(N-1)\mathbf{I}_{K}
\end{align}
and that $\mathbf{M}(\bar{\xi^{\ast}})$ has eigenvalues $\lambda^{\ast}_{1}=\dots=\lambda^{\ast}_{K}=N-1$. \par
Hence $\bar{\xi^{\ast}}$ is $\phi_{f}$-optimal ($A$- and $D$-optimal) in $\Xi_{(N,K,b)}$ \citep[e.g.,~see][Theorem~3.3]{jacroux2013note}.
%$\textbf{Theorem 3.3}$ \citep[e.g., see][p.1164]{jacroux2013note}
\end{steps}

\section*{Method~15:}%$n=pk^2\equiv2$ $(\bmod \ 4)$}
For this method of constructing main effects paired comparison block design $\bar{\xi^{\ast}}$, let $\textbf{L}$ be a $p\times q$ Hadamard matrix with entries of the corresponding values of $\mathbf{A}$ and let $\Xi_{(N,qK,pm)}$ be the class of designs where $N=pm^2\equiv2$ $(\bmod \ 4)$ and $m=3$. The design $\bar{\xi^{\ast}}\in\Xi_{(pm^2,qK,pm)}$ is constructed as follows:
\begin{steps}
\item Let design $d_1\in\Xi_{(m^2,K,m)}$ having matrix $\textbf{F}_{d_1}$ be a design which is obtained through construction Method $14$. 

\item Now form the desired design $\bar{\xi^{\ast}}\in\Xi_{(pm^2,qK,pm)}$ having matrix
\begin{equation*}
\textbf{F}= \textbf{L}\otimes \textbf{F}_{d_1}.
\end{equation*}\vspace{3mm}\par
From \eqref{eq:4.56} it is not difficult to see that $\bar{\xi^{\ast}}$ has $$\mathbf{M}(\bar{\xi^{\ast}})=(pm^2-p)\mathbf{I}_{qK}$$ and that $\mathbf{M}(\bar{\xi^{\ast}})$ has maximal trace among all designs in $\Xi_{(pm^2,qK,pm)}$. Hence $\bar{\xi^{\ast}}$ is $\phi_{f}$-optimal ($A$- and $D$-optimal) in $\Xi_{(pm^2,qK,pm)}$ \citep[e.g.,~see][Theorem~3.5]{jacroux2013note}.
%is optimal among all designs in $D(pk^2,qm,pk)$ by $\textbf{Theorem 2.3}$ of \cite{jacroux2013note} by $\textbf{Theorem 3.5}$ \citep[e.g., see][p.1165]{jacroux2013note}.
\end{steps}

\section*{Method~16:}% $N=2p+i\equiv i$ $(\bmod \ 4), i=1,2$ \rm{or} 3.}
For this method of constructing main effects paired comparison block design $\bar{\xi^{\ast}}$ within classes $\Xi_{(2p+i,K;m_1+1,\dots,m_i+1,m_{i+1},\dots,m_b)}$ for $N=2p+i\equiv i$ $(\bmod \ 4), i=1,2$ \rm{or} $3$ having $b$ blocks and block sizes $m_1,\dots,m_b$, we note that the different $m_1,\dots,m_b$ are all even and $p$ is the order of a Hadamard matrix. With this in mind the construction is given in the following steps:
\begin{steps}
\item Denote by $\textbf{H}_{p}$ a $p\times p$ Hadamard matrix with its first row and column having all $1$'s and let $\textbf{H}_{K}$ be a matrix consisting of $K$ columns from $\textbf{H}_{p}$. Now let
\begin{equation*}
\textbf{F}_{d_1}=\textbf{H}_{K} \otimes\mathbf{A}_v
\end{equation*}
where the corresponding design $d_1\in \Xi_{(2p,K)}$ has succeeding two profiles which are complementary.%, i.e., each succeeding two profiles have different signs for each attribute. 
\item Form a new design $d_2\in \Xi_{(2p,K;m_1,\dots,m_b)}$ where succeeding two complementary profiles in $\textbf{F}_{d_1}$ are combined to form blocks of sizes $m_1,\dots,m_b$, respectively, i.e., blocks of $d_2$ consist of $m_i/2$ complementary two profiles.
\item For $i=1,2$ or $3$, add a profile of all $2$'s to block $m_j, j=1,\dots i$, to form the desired design $\bar{\xi^{\ast}}\in\Xi_{(2p+i,K;m_1+1,\dots,m_i+1,m_{i+1},\dots,m_b)}$ having corresponding design matrix $\textbf{F}$. \vspace{3mm}\par
With regard to the final design $\bar{\xi^{\ast}}\in\Xi_{(2p+i,K;m_1+1,\dots,m_i+1,m_{i+1},\dots,m_b)}$ and from \eqref{eq:4.47} we observe that $\bar{\xi^{\ast}}$ has information matrix of the form
\begin{equation}\label{eq:5.57}
\mathbf{M}(\bar{\xi^{\ast}})=(N-i)\mathbf{I}_{K}+i\mathbf{J}_{K}-\sum_{j=1}^{i}\begin{pmatrix} \frac{1}{m_j+1} \end{pmatrix}\mathbf{J}_{K}
\end{equation}
and that $\bar{\xi^{\ast}}$ has maximum eigenvalue $\lambda^{\ast}=N-i=2p$. \par
Hence, $\bar{\xi^{\ast}}$ is $E$-optimal in $\Xi_{(2p+i,K;m_1+1,\dots,m_i+1,m_{i+1},\dots,m_b)}$ \\\citep[e.g.,~see][Theorem~3.1]{jacroux2017optimality}.
%which is optimal by $\textbf{Theorem 3.1}$ \citep[e.g., see][p.~5]{jacroux2017optimality}
\end{steps}

Now we give an example of the method of construction given in Method $16$.
\begin{exmp}\label{egmth16}
\rm{Suppose an experimenter is interested in constructing a blocked design $\bar{\xi^{\ast}}$ in classes $\Xi_{(17,K;3,4,4,6)}$ having $K=4$ attributes and $N=17$ paired comparisons which are to be partitioned into $4$ blocks of sizes $3,4,4$ and $6$. Then by Example \ref{eg2} the corresponding design matrix $\textbf{F}$ can be obtained from the transposed array or matrix with replaced attribute levels below:}\vspace{2mm}
{\small
$\begin{array}{lllllllllllllllll}
(1,2)	(2,1)	(1,2)	(2,1)	(1,2)	(2,1)	(1,2)	(2,1)	(1,2)	(2,1)	(1,2)	(2,1)	(1,2)	(2,1)	(1,2)	(2,1)	(1,2)\\
(1,2)	(2,1)	(2,1)	(1,2)	(1,2)	(2,1)	(2,1)	(1,2)	(1,2)	(2,1)	(2,1)	(1,2)	(1,2)	(2,1)	(2,1)	(1,2)	(1,2)\\
(1,2)	(2,1)	(1,2)	(2,1)	(2,1)	(1,2)	(2,1)	(1,2)	(1,2)	(2,1)	(1,2)	(2,1)	(2,1)	(1,2)	(2,1)	(1,2)	(1,2)\\
(1,2)	(2,1)	(2,1)	(1,2)	(2,1)	(1,2)	(1,2)	(2,1)	(1,2)	(2,1)	(2,1)	(1,2)	(2,1)	(1,2)	(1,2)	(2,1)	(1,2)\\
\end{array}$
}
\end{exmp}

\section*{Method~17:}% $n=2p+3\equiv 3$ $(\bmod \ 4)$.}
For this method of constructing block design $\bar{\xi^{\ast}}$, we denote the class of corresponding designs as $\Xi_{(2p+3,K;m_1+1,m_2,\dots,m_b)}$ where $N=2p+3\equiv 3$ $(\bmod \ 4)$, $p$ is the order of a Hadamard matrix and having $b$ blocks of sizes $m_1+1,m_2,\dots,m_b$. Here the block sizes $m_1,\dots,m_b$ can be even and odd. Now the construction is presented in the following steps: 

\begin{steps}
\item Let $\textbf{H}_{K}$ be a matrix consisting of $K$ columns selected from a $p\times p$ Hadamard matrix $\textbf{H}_{p}$ with its first row and column having all $1$'s. Now let
\begin{equation*}
\textbf{F}_{\bar{d}}= \begin{pmatrix} \textbf{H}_{K} \\ \textbf{J}_{1K} \end{pmatrix}\otimes\mathbf{A}_v
\end{equation*}
where $\textbf{J}_{pq}$ denotes a $p\times q$ matrix of $1$'s. Observe that $\bar{d}\in \Xi{(2p+2,K)}$ has succeeding two profiles which are complementary. 
\item Form a new design $\bar{d}_1\in \Xi_{(2p+2,K;m_1,\dots,m_b)}$ where succeeding two profiles which are complementary in $\textbf{F}_{\bar{d}}$ are combined to form blocks of sizes $m_1,\dots,m_b$, respectively. Note that block $i$ of $\bar{d}_1$ consists of $m_i/2$ complementary two profiles.
\item Now add a profile of all positive values corresponding to $\mathbf{A}$ to block $1$ of $\bar{d}_1$ to form the desired design $\bar{\xi^{\ast}}\in\Xi_{(2p+3,K;m_1+1,m_2,\dots,m_b)}$ having corresponding design matrix $\textbf{F}$. \vspace{3mm}\par

We observe that the design $\bar{\xi^{\ast}}\in\Xi_{(2p+3,K;m_1+1,m_2,\dots,m_b)}$ in view of \eqref{eq:4.47} has
\begin{equation}\label{eq:5.58}
\mathbf{M}(\bar{\xi^{\ast}})=2p\mathbf{I}_{K}+3\mathbf{J}_{K}-\frac{1}{m_1+1}\mathbf{J}_{K}
\end{equation}
and that $\bar{\xi^{\ast}}$ has maximum eigenvalue $\lambda^{\ast}=2p$ and is $E$-optimal in the corresponding classes $\Xi_{(2p+3,K;m_1+1,m_2,\dots,m_b)}$\\ \citep[e.g.,~see][Theorem~3.1]{jacroux2017optimality}.
%$\textbf{Theorem 3.1}$ \citep[e.g., see][p.~7]{jacroux2017optimality}
\end{steps}

\section*{Method~18:}% $n=pk_1+i\equiv i$ $(\bmod \ 4), i=1,2$ \rm{or} 3.}
Let $\textbf{H}_{p}$ be a $p\times p$ Hadamard matrix with its first row and column having all $+1$'s. Denote the class of available designs as $\Xi_{(pm_1+i,K;m_1+i,p-1-m_1-i;m_1+1,m_1)}$ where $N=pm_1+i\equiv i$ $(\bmod \ 4), i=1,2$ \rm{or} $3$ and having blocks $m_1+i$ and $p-1-m_1-i$ of sizes $m_1+1$ and $m_1$, respectively. Now let $\bar{\xi^{\ast}}$ denote the desired main effects paired comparison block designs in the corresponding classes $\Xi_{(pm_1+i,K;m_1+i,p-1-m_1-i;m_1+1,m_1)}$. Now the construction is presented in the following steps: 
\begin{steps}
\item  Let $\textbf{H}_{K}$ deonte the matrix consisting of $K$ columns from $\textbf{H}_{p}$. Now let $\boldsymbol{\iota}$ denote a $m_1\times 1$ vector containing $m_1/2$ positive values and $m_1/2$ negative values corresponding to $\mathbf{A}$ and let
\begin{equation*}
\textbf{F}_{d_1}=\textbf{H}_{K} \otimes\boldsymbol{\iota},
\end{equation*}
where the corresponding design $d_1\in \Xi_{(pm_1,K;p;m_1)}$ has succeeding sets of $m_1$ profiles in $\textbf{F}_{d_1}$ forming blocks. Also observe that block $1$ of $d_1$ consists of profiles of $\textbf{F}_{d_1}$ which contain all negative or positive values corresponding to $\mathbf{A}$.
\item Take each of the first $m_1$ profiles of all negative and positive values in $\textbf{F}_{d_1}$ and assign each such profile to one of the succeeding $m_1$ blocks of $d_1$ to form design $d_2\in \Xi_{(pm_1,K;m_1,p-1-m_1;m_1+1,m_1)}$.
\item For $i=1,2$ or $3$, take $i$ additional rows of all positive values corresponding to $\mathbf{A}$ and add each such row to a block of $d_2$ of size $m_1$ to form the final design $\bar{\xi^{\ast}}\in\Xi_{(pm_1+i,K;m_1+i,p-1-m_1-i;m_1+1,m_1)}$ having design matrix $\textbf{F}$.\vspace{3mm}\par

From \eqref{eq:4.47} we observe that the final design \\$\bar{\xi^{\ast}}\in\Xi_{(pk_1+i,K;m_1+i,p-1-m_1-i;m_1+1,m_1)}$ has information matrix $\mathbf{M}(\bar{\xi^{\ast}})$ of the form
\begin{equation}\label{eq:5.59}
\mathbf{M}(\bar{\xi^{\ast}})=pm_1\mathbf{I}_{K}+i\mathbf{J}_{K}-\frac{m_1+i}{m_1+1}\mathbf{J}_{K},
\end{equation}
which has maximum eigenvalue $\lambda^{\ast}=pm_1$ and hence $\bar{\xi^{\ast}}$ is $E$-optimal in the corresponding class of designs $\Xi_{(pk_1+i,K;m_1+i,p-1-m_1-i;m_1+1,m_1)}$ \citep[e.g.,~see][Theorem~3.1]{jacroux2017optimality}.
%$\textbf{Theorem 3.1}$ \citep[e.g., see][p.~8]{jacroux2017optimality} 
\end{steps}

\section*{Method~19:}% $bk=n\equiv 6$ $(\bmod \ 8),$ \textrm{$b$ $odd,$ $k$ $even.$}}
For this method of constructing main effects paired comparison block designs $\bar{\xi^{\ast}}$ within classes $\Xi_{(N,K,b)}$ having $b$ blocks (odd) and $m$ block sizes (even) where $bm=N\equiv 6$ $(\bmod \ 8)$, it is worth-while mentioning that if $m\equiv 2$ $(\bmod \ 8)$ the corresponding block is either $b\equiv 3$ $(\bmod \ 8)$ or $b\equiv 7$ $(\bmod \ 8)$ and if $m\equiv 6$ $(\bmod \ 8)$ the corresponding block is either $b\equiv 1$ $(\bmod \ 8)$ or $b\equiv 5$ $(\bmod \ 8)$. With this in mind now let $b=2+(b-2)$, $N=N_1+N_2$ for which run $N_1=2m\equiv 4$ $(\bmod \ 8)$ as well as run $N_2=m(b-2)\equiv 2$ $(\bmod \ 8)$ and the $K$ attributes satisfying $K\leq \min\{ 2m-2, (N_2-2)/2\}$. The construction is given in the following steps: 
\begin{steps}
\item Let $\textbf{H}_{2m}$ be a $2m\times 2m$ Hadamard matrix with its first row and column having all $1$'s. Now select column two of $\textbf{H}_{2m}$ as a blocking column and form two blocks of size $m$ where one block consists of rows with $+1$ in column two and one block consists of rows with $-$1 in column two. Arrange the rows of $\textbf{H}_{2m}$ so that one block consists of the first $m$ rows and the other block consists of the remaining $m$ rows. Delete columns one and two from $\textbf{H}_{2m}$, and let $\textbf{H}_1$ denote the resulting matrix with corresponding attributes $K$. Now let the design $d_1\in \Xi_{(N_1, K, b_1)}$ have $\textbf{F}_{d_1}= \textbf{H}_{1}\cdot \mathbf{A}$.

\item Let $\textbf{H}_{2}$ be a Hadamard matrix of order $(N_2-2)/2\times (N_2-2)/2$ and let
\begin{equation*}
\textbf{L}_{2}= \begin{pmatrix} \textbf{H}_{2} \\ \textbf{J}_{1, (N_2-2)/2} \end{pmatrix}.
\end{equation*}
\item Now let the design $d_2\in\Xi_{(N_2, K, b_2)}$ have matrix $\textbf{F}_{d_2}=\textbf{L}_{2}\otimes\mathbf{A}_v$ which is obtained by selecting $K$ columns from $\textbf{L}_{2}$ and where blocks are formed by combining $m/2$ successive two complementary profiles in $\textbf{F}_{d_2}$.
\item The desired design $\bar{\xi^{\ast}}\in\Xi_{(N,K,b)}$ has design matrix
\begin{equation*}
\textbf{F}= \begin{pmatrix} \textbf{F}_{d_1} \\ \textbf{F}_{d_2} \end{pmatrix}.
\end{equation*}  \vspace{3mm}\par
We observe that the final paired comparison block designs $\bar{\xi^{\ast}}\in\Xi_{(N,K,b)}$ has information matrix $\mathbf{M}(\bar{\xi^{\ast}})$ of the same form as \eqref{eq:4.49} and is hence $E$-optimal in $\Xi_{(N,K,b)}$ \citep[e.g.,~see][Corollary~3.2]{jacroux2015optimality}.
%which is optimal by $\textbf{Corollary 3.2}$ \citep[e.g., see][p.~166--167]{jacroux2015optimality}
\end{steps}

\section*{Method~20:}%$bk=n\equiv 2$ $(\bmod \ 8),$ \textrm{$b$ $even,$ $b\leq2k$ $and$ $k$ $odd,$  $k\geq4.$}}
For this method of construction we first note that the block $b\leq2m$ is even and the block size $m\geq4$ is odd. Now since $m$ is odd, $2m\equiv 2$ $(\bmod \ 8)$ with corresponding block $b\equiv 2$ $(\bmod \ 8)$ or $2m\equiv 6$ $(\bmod \ 8)$ with corresponding block $b\equiv 6$ $(\bmod \ 8)$. So now we denote the desired paired comparison block designs as $\bar{\xi^{\ast}}$ and let $\Xi_{(N, K, b)}$ denote the corresponding class of designs having pairs $N=N_1+N_2$ for $bm=N\equiv 2$ $(\bmod \ 8)$ where $N_1=(b-2)(m-1)$, $N_2=b-2+2m\equiv 2$ $(\bmod \ 8)$ and $K$ attributes satisfying $K\leq \min\{ 2(m-2), (N_2-2)/2\}$. To implement the desired design $\bar{\xi^{\ast}}\in\Xi_{(N,K,b)}$, we adopt the following steps:
\begin{steps}
\item Denote by $\textbf{H}_{1}$ an $2(m-1)\times 2(m-1)$ Hadamard matrix having its first row all $1$'s and first column all $1$'s. Now use column two of $\textbf{H}_{1}$ as a blocking column and form two blocks with one block consisting of those runs with $1$ in column two and the second block consisting of those runs with $-1$ in column two. Arrange the runs of $\textbf{H}_{1}$ accordinly so that block one occupies the first $M$ $-1$ runs of $\textbf{H}_{1}$. Now move the runs off all $1$'s in $\textbf{H}_{1}$ to the opposite block so that $\textbf{H}_{1}$ now has two blocks, one having $m-2$ runs with a total of $-1$ in each column of the block and one having $m$ runs with a total of $+1$ in each column of the block. Suppress columns one and two from $\textbf{H}_{1}$, and let $\textbf{H}_{2}$ denote the resulting matrix with corresponding attributes $K$. Now, let the design $\bar{d}_1$ have design matrix $\textbf{F}_{\bar{d}_1}= (\textbf{J}_{(b-2)/2,1}\otimes\textbf{H}_{2})\cdot\mathbf{A}$, and observe that $\bar{d}_1$ has $(b-2)/2$ blocks of size $m-2$ and $(b-2)/2$ blocks of size $m$.

\item Let $\textbf{H}_{3}$ be an $(N_2-2)/2\times (N_2-2)/2$  Hadamard matrix and let
\begin{equation*}
\textbf{L}_{3}= \begin{pmatrix} \textbf{H}_{3} \\ \textbf{J}_{1, (N_2-2)/2} \end{pmatrix}\otimes\mathbf{A}_v
\end{equation*}
Now let $\bar{d}_2$ have matrix $\textbf{F}_{\bar{d}_2}$ which is obtained by selecting $K$ columns from $\textbf{L}_{3}$.
\item Finally, we obtain the desired design $\bar{\xi^{\ast}}\in\Xi_{(N,K,b)}$ by adding the profiles of $\textbf{F}_{\bar{d}_2}$ to $\textbf{F}_{\bar{d}_1}$. Specifically, form the design $\bar{\xi^{\ast}}$ having design matrix $\textbf{F}$, which is obtained from $\textbf{F}_{\bar{d}_1}$ by first adding succeeding complementary two profiles from the $\textbf{H}_{3}\otimes\mathbf{A}_v$ portion of $\textbf{F}_{\bar{d}_2}$ to each block of size $m-2$ in $\textbf{F}_{\bar{d}_1}$. Now observe that $\textbf{F}$ has $b-2$ blocks of size $m$ and $\textbf{F}_{\bar{d}_2}$ has $m/2$ remaining complementary two profiles including two profiles with one profile having all $2$'s and one profile having all $-2$'s. Now form the last two blocks of $\textbf{F}$ by using $(m-1)/2$ of the remaining complementary two profiles of $\textbf{F}_{\bar{d}_2}$ along with the remaining profile of all $2$'s in $\textbf{F}_{\bar{d}_2}$ to form one block and using $(m-1)/2$ of the remaining complementary two profiles in $\textbf{F}_{\bar{d}_2}$ along with the remaining run of all $-2$'s in $\textbf{F}_{\bar{d}_2}$ to form the other block, which results in the design $\bar{\xi^{\ast}}$.\vspace{3mm}\par

Now from \eqref{eq:4.47} we observe that the final design $\bar{\xi^{\ast}}\in\Xi_{(N,K,b)}$ obtained in the construction process has information matrix 
\begin{equation}\label{eq:4.60}
\mathbf{M}(\bar{\xi^{\ast}})= (1/N)\Big(\mathbf{F}^{\top}\mathbf{F}-(1/m)\mathbf{F}^{\top}\mathbf{Z}\mathbf{Z}^{\top}\mathbf{F} \Big)=(N-2)\mathbf{I}_{K}+2\mathbf{J}_{K}-(b/m)\mathbf{J}_{K}
\end{equation}
and that $\mathbf{M}(\bar{\xi^{\ast}})$ has eigenvalues $\lambda^{\ast}_{1}=\dots=\lambda^{\ast}_{K-1}=N-2$ and $\lambda^{\ast}_{K}=N+2(K-1)-Kb/m$. \par
Hence, since $b\leq2m$ we obtain $\lambda^{\ast}_{1}\leq\lambda^{\ast}_{K}$, and that $\bar{\xi^{\ast}}$ is $E$-optimal in $\Xi_{(N,K,b)}$ \citep[e.g.,~see][Corollary 3.2]{jacroux2015optimality}.
%$\textbf{Corollary 3.2}$ \citep[e.g., see][p.~169--170]{jacroux2015optimality}
\end{steps}

\section*{Method~21:}% $bk=n\equiv 6$ $(\bmod \ 8),$ \textrm{$b$ $even,$ $b\leq2k$ $and$ $k$ $odd,$  $k\geq4.$}}
For this method of constructing main effects paired comparison block designs $\bar{\xi^{\ast}}$ within classes $\Xi_{(N,K,b)}$ where $bm=N\equiv 6$ $(\bmod \ 8)$, $b\leq2m$ for $b$ even and $m$ odd, we note that when $b\equiv 6$ $(\bmod \ 8)$, $m\equiv 5$ $(\bmod \ 8)$ or $m\equiv 1$ $(\bmod \ 8)$ and that when $b\equiv 2$ $(\bmod \ 8)$, $m\equiv 7$ $(\bmod \ 8)$ or $m\equiv 3$ $(\bmod \ 8)$. It follows that for $m$ odd and run $bm\equiv 6$ $(\bmod \ 8)$, we can always find $m_1$ (even) and $m_2$ (odd) such that $m_1+m_2=m$, $bm_1=N_1\equiv 4$ $(\bmod \ 8)$ and $bm_2=N_2\equiv 4$ $(\bmod \ 8)$ such that $N_1+N_2=N$. To satisfy the underlying construction process, we apply exhaustive search to obtain the values of $m_1$ and $m_2$ so that they maximize $\min\{ 2(m_1-1), (N_2-2)/2\}$ since the number of attributes $K$ must also satisfy $K\leq \min\{ 2(m_1-1), (N_2-2)/2\}$. The design construction is given in the following steps:
\begin{steps}
\item Let $\textbf{H}_{1}$ be a $2m_1\times 2m_1$ Hadamard matrix having its first row and column all $1$'s. Now use a single column of $\textbf{H}_{1}$, say the second column, as a blocking column and form two blocks with one block consisting of those runs with $1$ in column two and the second block consisting of those runs with $-1$ in column two. Arrange the runs of $\textbf{H}_{1}$ so that block one occupies the first $m_1$ runs of $\textbf{H}_{1}$. Now move the runs off all $1$'s in $\textbf{H}_{1}$ to the opposite block so that now $\textbf{H}_{1}$ has two blocks, one having $m_1-1$ runs and the other block has $m_1+1$ runs. Eliminate columns one and two from $\textbf{H}_{1}$, and let $\textbf{H}_{2}$ denote the resulting matrix with corresponding attributes $K$. Now let $d_1$ be the design having $\textbf{F}_{d_1}= (\textbf{J}_{(b/2),1}\otimes\textbf{H}_{2})\cdot\mathbf{A}$ and observe that $d_1$ has $b/2$ blocks of size $m_1-1$ and $b/2$ blocks of size $m_1+1$.

\item Let $\textbf{H}_{3}$ be an $(N_2-2)/2\times (N_2-2)/2$ Hadamard matrix and let
\begin{equation*}
\textbf{L}_{3}= \begin{pmatrix} \textbf{H}_{3} \\ \textbf{J}_{1, (N_2-2)/2} \end{pmatrix}\otimes\mathbf{A}_v
\end{equation*}
Now let $d_2$ be a design having $\textbf{F}_{d_2}$ which is obtained by selecting $K$ columns from $\textbf{L}_{3}$.
\item Finally, we obtain the desired design $\bar{\xi^{\ast}}\in\Xi_{(N,K,b)}$ having design matrix $\textbf{F}$ by using succeeding complementary two profiles from $\textbf{F}_{d_2}$. Specifically, $\textbf{F}$ is obtained from $\textbf{F}_{d_1}$ by adding $(m-m_1+1)/2$ complementary two profiles from $\textbf{F}_{d_2}$ to each block of $\textbf{F}_{d_1}$ having size $m_1-1$ and adding $(m-m_1-1)/2$ complementary two profiles from $\textbf{F}_{d_2}$ to each block of $\textbf{F}_{d_1}$ having size $k_1+1$. The resulting design $\bar{\xi^{\ast}}$ has $b$ blocks of size $m$ and is the desired design.  \vspace{3mm}\par
We observe that the final design $\bar{\xi^{\ast}}\in\Xi_{(N,K,b)}$ has $\mathbf{M}(\bar{\xi^{\ast}})$ of the form \eqref{eq:4.60} with $\lambda^{\ast}_{1}=N-2$ since $b\leq2m$, and that $\bar{\xi^{\ast}}$ is $E$-optimal in $\Xi_{(N,K,b)}$ \citep[e.g.,~see][Corollary 3.2]{jacroux2015optimality}.
%$\textbf{Corollary 3.2}$ \citep[e.g., see][p.~171--172]{jacroux2015optimality}
\end{steps}

\section*{Method~22: }%$bk=n\equiv 1$ $(\bmod \ 4),$ \textrm{$b$ $odd,$ $k$ $odd,$ $b\leq k.$}}
For this method of constructing main effects paired comparison block designs $\bar{\xi^{\ast}}$ within classes $\Xi_{(N,K,b)}$ where $bm=N\equiv 1$ $(\bmod \ 4)$ for both $b$ and $m$ odd and $b\leq m$, we note that if the block $b\equiv 1$ $(\bmod \ 4)$ the corresponding block size is equal to $m\equiv 1$ $(\bmod \ 4)$ or if the block $b\equiv 3$ $(\bmod \ 4)$  the corresponding block size is equal to $m\equiv 3$ $(\bmod \ 4)$. We further note that unlike Method $14$ where classes having $N\equiv 1$ $(\bmod \ 4)$ and $b=m$ were considered here we focus on more general classes having $N\equiv 1$ $(\bmod \ 4)$ and $b\leq m$ with a different optimality criterion. Now the construction is given in the following steps:
\begin{steps}
\item Select even integers $q_i$, $i=1,\dots, (b-3)/2$, so that each $q_i\leq m$ and that
 \begin{equation*}
 \begin{split}
&2\{(m-q_1)+\dots +(m-q_{(b-3)/2})\}+(3m-1)  \\
\qquad&=bm-1-2(q_1+\dots +q_{(b-3)/2}) \\
\qquad&=(N-1)-2(q_1+\dots +q_{(b-3)/2})  \\
&=\textbf{M}\equiv 0\ (\bmod \ 8).
\end{split}
\end{equation*}
We note that $q_i$ should be selected so that $\textbf{P}=\min\{2q_1-2,\dots,2q_{(b-3)/2}-2, \textbf{M}/2\}$ is as large as possible.
\item For each $q_i$, select a $2q_i\times 2q_i$ Hadamard matrix $\textbf{H}_{i}$ which has all 1's in its first row and column, $i=1,\dots,(b-3)/2$. For each $\textbf{H}_{i}$, select one of the columns having $q_i$ $1$'s and $q_i$ $-1$'s in it as a blocking column and put all the rows having a $1$ in the blocking column in one block and the remaining rows in the second block. Now drop the first column and the blocking column from each $\textbf{H}_{i}$ and select $K$ of the remaining columns from each $\textbf{H}_{i}$ so that $K\leq \textbf{P}$ where $\textbf{P}$ is defined as in step 1 and let the resulting matrix be denoted by $\textbf{L}_{i}$. We note that each $\textbf{L}_{i}$ is a $2q_i\times K$ matrix consisting of two blocks of size $q_i$. Finally, move the row of all 1's in each $\textbf{L}_{i}$ from one block to the other and let the resulting matrix be denoted by $\textbf{T}_{i}$. We note that $\textbf{T}_{i}$ is a $2q_i\times K$ matrix having two blocks with one block of size $q_i-1$ and the other of size $q_i+1$. We also note that all column sums in the block of size $q_i-1$ are equal to $-1$ and all column sums in the block of size $q_i+1$ are equal to $1$.   
\item Now select an $(\textbf{M}/2)\times (\textbf{M}/2)$ Hadamard matrix $\textbf{H}_{b}$ having its first row and column all $1$'s and let 
\begin{equation*}
\textbf{L}_{b}= \textbf{H}_{b}\otimes\mathbf{A}_v.
\end{equation*}
Select $K$ columns from $\textbf{L}_{b}$ and let the resulting matrix be denoted by $\textbf{L}_{c}$. We note that $\textbf{L}_{c}$ is an $\textbf{M}\times K$ matrix and that each succeeding two profiles in $\textbf{L}_{c}$ are complementary, i.e., each two profiles in $\textbf{L}_{c}$ have columns sums zero in all $K$ columns.
\item For $i=1,\dots,(b-3)/2$ and $\textbf{T}_{i}$ obtained in step 2, take $(m-q_i+1)/2$ complementary two profiles from $\textbf{L}_{c}$ (excluding the profiles of all $2$'s and $-1$'s in $\textbf{L}_{c}$) and add them to the block of $\textbf{T}_{i}$ of size $q_i-1$ and take $(m-q_i+1)/2$ complementary two profiles from $\textbf{L}_{c}$ (excluding the profiles of all $2$'s and all $-2$'s in $\textbf{L}_{c}$) and add them to the block of $\textbf{T}_{i}$ of size $q_i+1$ and let the resulting matrix be denoted by $\textbf{F}_{di}$, $i=1,\dots,(b-3)/2$. We note that each $\textbf{F}_{di}$ is a $2m\times K$ design matrix having two blocks of size $m$ with all column sums in one block equal to $-2$ and all column sums in other block equal to $2$.
\item Now form the $3m\times K$ design matrix $\textbf{F}_{db}$ which consists of three blocks of size $m$ where the first block consists of $(m-1)/2$ complementary two profiles along with the profile of all $2$'s from $\textbf{L}_{c}$, the second block consists of $(m-1)/2$ complementary two profiles along with the profile of all $-2$'s from $\textbf{L}_{c}$ and the third block consists of $(m-1)/2$ complementary profiles from $\textbf{L}_{c}$ along with the profile of all $2$'s which is added. Thus $\textbf{F}_{db}$ consists of three blocks of size $m$ with the column sums of all attributes in two of the blocks equal to $2$ and the column sums of all attributes in the remaining block equal to $-2$.
\item Obtained the desired design $\bar{\xi^{\ast}}$ within classes $\Xi_{(N,K,b)}$ having design matrix
\begin{equation*}
\textbf{F}^{\top}= \begin{pmatrix} \textbf{F}^{\top}_{d1}, \dots, \textbf{F}^{\top}_{d,(b-3)/2}, \textbf{F}^{\top}_{db}\end{pmatrix}.
\end{equation*}
For the design $\bar{\xi^{\ast}}\in\Xi_{(N,K,b)}$ obtained using this construction process, all attributes in $(b-1)/2$ of the blocks have columns sums $-2$ and all attributes in $(b+1)/2$ of the blocks have column sums $2$. \vspace{3mm}\par
Now we observe that if the profiles of the design matrix $\textbf{F}$ are assigned to blocks as specified by the block indicator matrix $\mathbf{Z} = \mathbf{I}_{b}\otimes \mathbf{1}_{m}$ then the final design $\bar{\xi^{\ast}}\in\Xi_{(N,K,b)}$ obtained in the construction process has information matrix 
\begin{equation}\label{eq:4.61}
\mathbf{M}(\bar{\xi^{\ast}})= \mathbf{F}^{\top}\mathbf{F}-(1/m)\mathbf{F}^{\top}\mathbf{Z}\mathbf{Z}^{\top}\mathbf{F} =(N-1)\mathbf{I}_{K}+\mathbf{J}_{K}-(b/m)\mathbf{J}_{K}
\end{equation}
and that $\mathbf{M}(\bar{\xi^{\ast}})$ has eigenvalues $\lambda^{\ast}_{1}=\dots=\lambda^{\ast}_{K-1}=N-1$ and $\lambda^{\ast}_{K}=(N-1)+K(1-b/m)$. \par
Hence, $\bar{\xi^{\ast}}$ is $E$-optimal in $\Xi_{(N,K,b)}$ \citep[e.g.,~see][Theorem~3.3]{jacroux2016optimality}.
%$\textbf{Theorem 3.3}$ of \citet{jacroux2016optimality} 
\end{steps}

\section*{Method~23:}%$n\equiv2$ $(\bmod \ 4)$ \rm{having different even block sizes where $m\geq 3$}.}
For this method of main effects paired comparison block designs $\bar{\xi^{\ast}}$ construction, we use similar notations such as $\Xi_{(N,K,b,m_1,m_2,\dots,m_b)}$ to denote a class of designs where $N\equiv2$ $(\bmod \ 4)$ and the matrix $\textbf{H}_{\frac{n}{2}-1}$ as already defined in Method $11$. To obtain the desired design $\bar{\xi^{\ast}}\in\Xi_{(N,K,b,m_1,m_2,\dots,m_b)}$, we use the following steps:
\begin{steps}
\item Let $\textbf{L}$ be the $(n-2)\times \frac{n}{2}-1$ matrix given by
\begin{equation*}
\textbf{L}=\textbf{H}_{\frac{n}{2}-1}\otimes\mathbf{A}_v.
\end{equation*}
\item Now augment $\textbf{L}$ by one profile of all $2$'s and another profile with $K_1$ entries equal to $2$ and $K_2$ entries equal to $-2$ where attribute $K=K_1+K_2=\frac{n}{2}-1$ and $K_2\geq 1$.
\item Obtain $\bar{\xi^{\ast}}$ having design matrix $\textbf{F}$ formed from augmented $\textbf{L}$ by selecting any $K$ columns with at least one column having the last two entries equal to $2$ and $-2$ where sets of successive $m_i$ profiles correspond to the $i$th block of $\bar{\xi^{\ast}}$, $i=1,2,\dots,b$. \vspace{3mm}\par
We observe that $\bar{\xi^{\ast}}$ has information matrix $\mathbf{M}(\bar{\xi^{\ast}})$ of the same form as \eqref{eq:4.49} and is hence $E$-optimal \citep[e.g., see][Theorem~3.5]{saharay2018optimality}.
%$\textbf{Theorem 3.5}$ \citep[e.g., see][p.~145]{saharay2018optimality}
\end{steps}

\section*{Method~24:}% $n\equiv2$ $(\bmod \ 4)$ \rm{where $m\geq 3$ and the block sizes can be even or odd}.}
For this method of constructing block design $\bar{\xi^{\ast}}$, we adopt the notations in Method $11$ and analogously denote $\Xi_{(N,K,b,m_1,m_2,\dots,m_b)}$ as a class of designs where $N\equiv2$ $(\bmod \ 4)$ and $K\geq 3$. Here the block sizes $m_1,m_2,\dots,m_b$ can be a mixture of even or odd and not necessarily equal. Now for the design $\bar{\xi^{\ast}}\in\Xi_{(N,K,b,m_1,m_2,\dots,m_b)}$, let the set of $x$ be blocks of even size, $y$ be blocks of size $3$ $(\bmod \ 4)$ and $z$ be blocks of size $1$ $(\bmod \ 4)$, $x,y,z\geq 0$ be arranged in a lexicographical order where for integers $t_1$ and $t$, $m_1=m_2=2t_1$, with $\sum_{i=3}^{x}m_i = 8t+\ell$ for which $\ell\in\{0,2,4,6\}$ satisfying $\ell+3y+z\equiv2\ (\bmod \ 4)$, $\ell+z=y+2$. For notational convienience, in the underlying construction a matrix with a deleted first column and a deleted first two columns will be decorated with a single and double asterisk, respectively. The construction is given in the following steps:
\begin{steps}
\item Assume $t_1$, $t>0$ and let the Hadamard matrices $\textbf{H}_{4t_1}$ and $\textbf{H}_{4t}$ exist. Let the matrix
\begin{equation}\label{eq:4.62}
\textbf{L}_1=\begin{pmatrix} \textbf{H}_{4t}\\ \textbf{J}_{\frac{\ell}{2}\times4t}\end{pmatrix}\otimes\textbf{A}_v.
\end{equation}
Now, in the case when $t=0$, we reformulate $\textbf{L}_1$ in \eqref{eq:4.62} as
\begin{equation*}
\textbf{L}_1=\textbf{J}_{\frac{\ell}{2}\times4t_1-2}\otimes\textbf{A}_v,
\end{equation*}
and in the case when $t_1=0$ and $t=0$, we reformulate $\textbf{L}_1$ in \eqref{eq:4.62} as
\begin{equation*}
\textbf{L}_1=\textbf{J}_{\frac{\ell}{2}\times u}\otimes\textbf{A}_v,
\end{equation*}
where \\
$u=\min\{m_{x+1}, m_{x+2},\dots, m_{x+y}, m_{x+y+1}-2, m_{x+y+2}-2, \dots, m_{x+y+z}-2\}$.
\item (In case $y$ is a positive integer).\\
Let for $K=x+i$, $i=1, 2,\dots, y$, the matrix $\textbf{H}_{m_{K}+1}$ exist and let
\begin{equation}\label{eq:4.63}
\textbf{B}_i=\textbf{H}^*_{m_{K}+1}\cdot\textbf{A}; i=1,2,\dots,y,
\end{equation}
\item (In case $z$ is a positive integer).\\
Let for $s=x+y+j$, $j=1, 2,\dots, z$, $\textbf{H}_{m_{s}-1}$ exist and let
\begin{equation}\label{eq:4.64}
\textbf{C}_j=\begin{pmatrix} \textbf{1}^{\top}\\\textbf{H}^*_{m_{s}-1}\\ \textbf{1}^{\top}\end{pmatrix}\cdot\textbf{A}, j=1,2,\dots,z.
\end{equation}
\item Let $\textbf{L}_{11}, \textbf{L}_{12}, \textbf{B}_{i1}$, and $\textbf{C}_{j1}$ consists of $K$ columns of $\textbf{H}^{**}_{4t_1}, \textbf{L}_{1}, \textbf{B}_{i}$ and $\textbf{C}_{j}$, respectively, where $K\leq\min\{2(2t_1-1), 4t, m_{x+1}, m_{x+2},\dots, m_{x+y}, m_{x+y+1}-2, m_{x+y+2}-2, \dots, m_{x+y+z}-2\}$, ignoring $2(2t_1-1)$, and (or) $4t$ in case $t_1$ and (or) $t$ is zero. \par
Obtain the desired design $\bar{\xi^{\ast}}$ having matrix $\textbf{F}^{\top}=(\textbf{L}^{\top}_{11}, \textbf{L}^{\top}_{12}, \textbf{B}^{\top}_{11},\dots,\textbf{B}^{\top}_{y1},$ $\textbf{C}^{\top}_{11},\dots,\textbf{C}^{\top}_{z1})$, where the sets of successive $m_i$ profiles correspond to the $i$th block of $\bar{\xi^{\ast}}$, $i=1,2,\dots b$.
\vspace{3mm}\par
We observe that if the profiles of the design matrix $\textbf{F}$ are assigned to blocks as specified by the block indicator matrix $\mathbf{Z} = \mathbf{I}_{b}\otimes \mathbf{1}_{m_i}$ then the final design $\bar{\xi^{\ast}}\in\Xi_{(N,K,b,m_1,m_2,\dots,m_b)}$ obtained in the construction process has information matrix 
\begin{equation}\label{eq:4.65}
\mathbf{M}(\bar{\xi^{\ast}})= \mathbf{F}^{\top}\mathbf{F}-(1/m_i)\mathbf{F}^{\top}\mathbf{Z}\mathbf{Z}^{\top}\mathbf{F} =(N-2)\mathbf{I}_{K}+(2-\delta)\mathbf{J}_{K}
\end{equation}
where the quantity $1<\delta<2$ and hence $\bar{\xi^{\ast}}$ is $D$-optimal in its corresponding classes $\Xi_{(N,K,b,m_1,m_2,\dots,m_b)}$ \citep[e.g.,~see][Theorem~3.4]{saharay2018optimality}.

%The construction of the desired design $d^{*}$ which is optimal in $\bar{D}(n,m,b,k_1,k_2,\dots,k_b)$ by $\textbf{Theorem 3.4}$ \citep[e.g., see][p.~145--146]{saharay2018optimality}
\end{steps}

\section*{Method~25:}% $n\equiv2$ $(\bmod \ 4)$ \rm{where $m\geq 3$ and the block sizes can be even or odd}.}
By adopting similar notations in Method $24$, we can also construct design $\bar{\xi^{\ast}}$ within classes $\Xi_{(N,K,b,m_1,m_2,\dots,m_b)}$. Here $m_1=4t_1+\ell_{1}$, with $\sum_{i=2}^{x}m_i = 8t+\ell_{2}$, where $\ell_{1}, \ell_{2}\in\{0,2,4,6\}$ satisfying $\ell_{1}+\ell_{2}+3y+z\equiv2\ (\bmod \ 4)$, $\ell_{1}+\ell_{2}+z=y+2$.  Now the construction is given in the following steps:
\begin{steps}
\item Assume $t_1$, $t>0$. \\
Let $\textbf{H}_{4t_1}$ and $\textbf{H}_{4t}$ exist. Let matrix
\begin{equation}\label{eq:4.66}
\textbf{L}_0=\begin{pmatrix} \textbf{1}^{\top}\\ \textbf{H}^{*}_{4t_1}\\ \textbf{J}_{\frac{\ell_1}{2}\times4t_1-1}\\ -\textbf{J}_{\frac{\ell_1}{2}\times4t_1-1}\end{pmatrix}\cdot\textbf{A},
\end{equation}
and
\begin{equation}\label{eq:4.67}
\textbf{L}_1=\begin{pmatrix}\textbf{H}_{4t}\\ \textbf{J}_{\frac{\ell_2}{2}\times4t}\end{pmatrix}\otimes\textbf{A}_v.
\end{equation}
Hence, in the case when $t=0$, we reformulate \eqref{eq:4.67} as follows:
\begin{equation*}
\textbf{L}_1=\textbf{J}_{\frac{\ell_2}{2}\times4t_1-1}\otimes\textbf{A}_v,
\end{equation*}
and in the case $t_1=0$, we reformulate \eqref{eq:4.66} as follows:
\begin{equation*}
\textbf{L}_0=\textbf{J}_{\frac{\ell_1}{2}\times 4t}\otimes\textbf{A}_v.
\end{equation*}
Now when $t_1=0$ and $t=0$, we reformulate \eqref{eq:4.66} and \eqref{eq:4.67} as follows:
\begin{equation*}
\textbf{L}_0=\textbf{J}_{\frac{\ell_1}{2}\times u}\otimes\textbf{A}_v,
\end{equation*}
\begin{equation*}
\textbf{L}_1=\textbf{J}_{\frac{\ell_2}{2}\times u}\otimes\textbf{A}_v,
\end{equation*}
where \\
$u=\min\{m_{x+1}, m_{x+2},\dots, m_{x+y}, m_{x+y+1}-2, m_{x+y+2}-2, \dots, m_{x+y+z}-2\}$.
\item (In case $y$ is a positive integer).\\
Let for $K=x+i$, $i=1, 2,\dots, y$, matrix $\textbf{H}_{m_{K}+1}$ exist and let
\begin{equation*}
\textbf{B}_i=\textbf{H}^*_{m_{K}+1}\cdot\textbf{A}; i=1,2,\dots,y.
\end{equation*}
\item (In case $z$ is a positive integer).\\
Let for $s=x+y+j$, $j=1, 2,\dots, z$, $\textbf{H}_{K_{s}-1}$ exist and let matrix
\begin{equation*}
\textbf{C}_j=\begin{pmatrix} \textbf{1}^{\top}\\\textbf{H}_{k_{s}-1}\\ \textbf{1}^{\top}\end{pmatrix}\cdot\textbf{A}, j=1,2,\dots,z.
\end{equation*}
\item Let $\textbf{L}_{10}, \textbf{L}_{11}, \textbf{B}_{i1}$, and $\textbf{C}_{j1}$ consist of $K$ columns of $\textbf{L}_{0}, \textbf{L}_{1}, \textbf{B}_{i}$ and $\textbf{C}_{j}$, respectively, where $K\leq\min\{4t_1-1), 4t, m_{x+1}, m_{x+2},\dots, m_{x+y}, m_{x+y+1}-2, m_{x+y+2}-2, \dots,$  $m_{x+y+z}-2\}$, ignoring $(4t_1-1)$, and (or) $4t$ in case $t_1$ and (or) $t$ is zero. \par
\item Now obtain the final design $\bar{\xi^{\ast}}\in\Xi_{(N,K,b,m_1,m_2,\dots,m_b)}$ having matrix $\textbf{F}=(\textbf{L}^{\top}_{10}, \textbf{L}^{\top}_{11},$ $ \textbf{B}^{\top}_{11},\dots,\textbf{B}^{\top}_{y1},\textbf{C}^{\top}_{11},\dots,\textbf{C}^{\top}_{z1})$, where the sets of successive $m_i$ profiles correspond to the $i$th block of $\bar{\xi^{\ast}}$, $i=1,2,\dots b$.\vspace{3mm}\par

We observe that the final design $\bar{\xi^{\ast}}$ obtained in the construction process has the same information matrix $\mathbf{M}(\bar{\xi^{\ast}})$ as in \eqref{eq:4.65} and hence $\bar{\xi^{\ast}}$ is $D$-optimal in its corresponding classes $\Xi_{(N,K,b,m_1,m_2,\dots,m_b)}$ \citep[e.g.,~see][Theorem~3.4]{saharay2018optimality}.
\end{steps}

\section*{Method~26:}%. $n\equiv1$ $(\bmod \ 4)$ \rm{where $m\geq 3$ and the block sizes can be even or odd}.}
Following the steps of construction described in Method $24$ we can also construct an optimal blocked main effects paired comparison design $\bar{\xi^{\ast}}$ within a given classe $\Xi_{(N,K,b,m_1,m_2,\dots,m_b)}$ where $N\equiv1$ $(\bmod \ 4)$. Here we note that for integers $t_1$ and $t$, $m_1=m_2=2t_1$, with $\sum_{i=3}^{x}m_i = 8t+\ell$ for which $\ell\in\{0,2,4,6\}$ satisfying $\ell+3y+z\equiv1\ (\bmod \ 4)$, $\ell+z=y+1$. It immediately follows from \eqref{eq:4.65} that the desired design $\bar{\xi^{\ast}}$ obtained through Method $24$ has information matrix $\mathbf{M}(\bar{\xi^{\ast}})$ of the form
\begin{equation}\label{eq:4.68}
\mathbf{M}(\bar{\xi^{\ast}})= \mathbf{F}^{\top}\mathbf{F}-(1/m_i)\mathbf{F}^{\top}\mathbf{Z}\mathbf{Z}^{\top}\mathbf{F} =(N-1)\mathbf{I}_{K}+(1-\delta)\mathbf{J}_{K}
\end{equation}
where the quantity $\delta<1$, and hence $\bar{\xi^{\ast}}$ is $E$-optimal in $\Xi_{(N,K,b,m_1,m_2,\dots,m_b)}$ \citep[e.g.,~see][Theorem~3.3]{dutta2015d}.
 \newpage

\section*{Method~27:}% $n\equiv1$ $(\bmod \ 4)$ \rm{where $m\geq 3$ and the block sizes can be even or odd}.}
Accordingly, by following the construction Method $25$ we can also construct an optimal blocked main effects paired comparison design $\bar{\xi^{\ast}}$ in its corresponding classes $\Xi_{(N,K,b,m_1,m_2,\dots,m_b)}$ where again $N\equiv1$ $(\bmod \ 4)$. Here we note that for integers $t_1$ and $t$, $m_1=4t_1+\ell_{1}$, with $\sum_{i=2}^{x}m_i = 8t+\ell_{2}$, where $\ell_{1}, \ell_{2}\in\{0,2,4,6\}$ satisfying $\ell_{1}+\ell_{2}+3y+z\equiv1\ (\bmod \ 4)$, $\ell_{1}+\ell_{2}+z=y+1$. It immediately follows that the desired design $\bar{\xi^{\ast}}$ obtained via Method $25$ has the same information matrix $\mathbf{M}(\bar{\xi^{\ast}})$ as in \eqref{eq:4.68} and hence $\bar{\xi^{\ast}}$ is $E$-optimal in $\Xi_{(N,K,b,m_1,m_2,\dots,m_b)}$ \citep[e.g.,~see][Theorem~3.3]{dutta2015d}. \vspace{3mm}\\

It is worth noting that for attributes $3\leq K \leq 12$ each at two-levels \citet{singh2015optimal} provided constructions of optimal paired choice block designs, which can be obtained from Theorem $3.3$ and Corollary $3.3$ as exhibited in Table $2$ of \citet{singh2015optimal} using the method of orthogonal array ($OA$) of strength two and generators ($G$), and having $N\equiv0$ $(\bmod \ 4)$ pairs. In their $OA+G$ construction method when the number of blocks $b=1$ the block size is given by $m=N\equiv0$ $(\bmod \ 4)$ and for $b=2$ or $4$ the block size can be either $m\equiv2$ $(\bmod \ 4)$ or $m\equiv0$ $(\bmod \ 4)$. For example, in their Table $2$ of optimal blocked designs constructions, entries of the form $(m,b)$ for the factorial $2^K$, $3\leq K \leq 12$ are shown in the first column, say, SDC (2015) of Table \ref{tab4.9} as against the corresponding various methods, say, $M_i$, $i=1,\dots,27$ main effects paired comparison block designs constructions presented herein.  

\begin{landscape}
\begin{longtable}{p{1cm}p{2.3cm}p{2.4cm}p{2.5cm}p{2.5cm}p{2.5cm}p{2.5cm}p{2.5cm}p{2.3cm}p{2.5cm}}
\caption{Optimal paired comparison block designs for binary $K$ attributes\label{tab4.9}} \\
\toprule
 &SDC (2015)  & $M_{1,2}$     & $M_{3,4}$        &  $M_{5,6,7}$        &   $M_{8}$   &  $M_{9,10}$ &$M_{11,12,13}$ \\
    &      $(m,b)$        & $(m,b)$           & $(m,b)$   & $(m,b)$        &  $(m,b)$        & $(m_1,m_2;b_1,b_2)$      &   $(m_{i};b)$           \\
$K$        &                                       &                               &                      & $(m_1,m_2;b_1,b_2)$      &        &    &                       \\ \midrule\endfirsthead
\caption*{\tablename{} \ref{tab4.9} (continued)}\\
\toprule
         & $M_{14,15}$  &$M_{16,17,18}$& $M_{19,20,21}$     & $M_{22}$   & $M_{23,24,25}$            &  & $M_{26,27}$   \\
 $K$    &  $(m,b)$        &$(m_i;b)$            & $(m,b)$                & $(m,b)$        & $(m_i;b)$                        & & $(m_i;b)$       \\ \midrule\endhead
\bottomrule\endfoot
3             &(4,1)                               & (2,5)                      & (6,3)             &(3,4)         &          &(3,2;6,4)          &(2,4,4;3)                        \\\hline              
4            &(4,2),(8,1)                      & (2,5)                      & (6,3)             & (3,4)       &           &(3,2;6,4)           &(2,4,4;3)                       \\\hline              
5-6        &  (4,2),(8,1)                    &(2,9)                       & (6,3),(14,3)  & (3,8)       &           &(3,2;6,4)           &(2,4,4;3)                   \\
             &  (6,2),(12,1)                  & (6,5)                      &   (6,7)            &               &            &                         &(4,4,6;3)                     \\\hline              
7           &  (8,1),(6,2)                    & (2,9)                      &(6,3)               &(3,8)       &(9,3)    &(3,4;12,8)        &(4,4,6;3)                 \\
            &  (12,1),(4,4)                  &(6,5)                       &                      &               &              &(3,2;6,4)         &(2,4,4,4,4;5)                \\
             &  (8,2),(16,1)                  &                              &                        &              &              &                       &                                  \\   \hline              
8            &  (6,2),(12,1)                  &(2,9)                      &  (6,3)              &(3,8)      &(9,3)      &(3,4;12,8)      &(4,4,6;3)                 \\
            &  (4,4),(8,2)                   &(6,5)                       &(14,3)             &              &               &(3,2;6,4)        &(2,4,4,4,4;5)                 \\
            &  (16,1)                           &                               &  (6,7)             &              &               &                       &                                     \\   \hline              
9-10     &  (6,2),(12,1)                 & (2,13)                     &(14,3)            &(2,4;4,4) &(9,3)      &(3,2;6,4)         &(4,4,4,4,4,6;6)      \\
     &(4,4),(8,2)                    &                               & (6,7)            &                 &               &                         &(4,4,4,4,6;5)      \\
             & (16,1),(10,2)              &                                &                      &                &             &                        &(12,12,12,14;4)     \\
            &(20,1)                           &                               &                       &                 &             &                        &                                           \\  \hline   
11             &(12,1),(10,2)              & (2,13)                     & (6,7)              &(2,4;4,4)  & (9,3)    &(3,2;6,4)          &(4,4,4,4,4,6;6)                              \\
             &(20,1),(4,4)                &                              & (14,3)              &                 &             &                        &(4,4,4,4,6;5)        \\
        &(8,2),(16,1)                 &                               &                      &                  &             &                         &(6,6,6,8,8;5)   \\
            &(6,4),(12,2)                 &                              &                       &                 &             &                        &                             \\
            &(24,1)                         &                               &                      &                  &             &                        &                                        \\      \hline   
12            &(10,2),(20,1)              &   (2,13)                 & (6,7)              & (3,12)       &(9,3)    &(3,2;6,4)          &(12,12,12,14;4)   \\
            &(4,4),(8,2)                 &                               &(14,3)             & (2,4;4,4) &             &                        & (4,4,4,4,4,6;6)    \\
        &(16,1),(6,4)                  &                               &                      &(2,4;6,6)     &            &                         &  (4,4,4,4,6;5)       \\
         &(12,2),(24,1)              &                               &                      &                    &            &                          &(6,6,6,8,8;5)              \\
3    &(3,3)          &(3,2,2,2;4)   &                               & (5,5)   &                                                        (3,3,5,5,5,5;6) &                           &(4,5;2)      \\
      &                 &(5,6;2)          &                               &             &                                                        (4,3,3,3,3,5,5;7)&                        & (4,2,3;3)           \\
      &                    &              &                               &               &                                                          (4,4,3,3,3,3,5,5;8)&                     & (6,6,8,8,9;5)          \\\hline                
4    &(3,3)             &(3,2,2,2;4)       &                                & (5,5)   &                                                     (7,7,7,7,9,9,9,9,9,9;10)&           & (6,6,8,8,9;5)           \\
     &                      &(5,6;2)                  &                               & (9,5)    &                                                  (6,7,7,7,7,7,7,7,7,9,9,9,9;13)&  & (11,11,9,9,9;5)           \\\hline              
5-6 &(3,6)            &(3,4,4,6;4)&(6,9)                       & (5,5)   &                                                         (7,7,7,7,9,9,9,9,9,9;10)&            & (6,6,8,8,9;5)           \\
     &(3,18)           &(5,6,8;3)     & (11,6)                     &  (9,5)  &                                                       (6,7,7,7,7,7,7,7,7,9,9,9,9;13)&  & (11,11,9,9,9;5)           \\\hline              
7     & (3,6)              &(3,4,4,6;4)         & (9,5)                       &(9,5)   &                                              (7,7,7,7,9,9,9,9,9,9;10)&             &  (6,6,8,8,9;5)             \\
   &(3,18)             &(5,6,8;3)          & (11,6)                    &           &                                                     (6,7,7,7,7,7,7,7,7,9,9,9,9;13)&   &(11,11,9,9,9;5)    \\
     &                    &(3,2;3,4)           &                              &           &                                                        &                                                      & (6,6,6,6,6,7,7,9;8)            \\   \hline              
8    & (3,6)            &(3,4,4,6;4)         & (6,9)                     &(9,5)  &   (4,6,6,6,6,6;6)                               &                                                    &  (12,12,11,13,13;5)           \\
  & (3,18)         &(5,6,8;3)           & (11,6)                     &         & (6,6,6,8,8,8;6)                               &                                                     & (15,15,13,13,13;5)      \\
    &                  &(3,2;3,4)          &                                &        & (8,8,10,10,10,10,10;7)                   &                                                    &               \\   \hline              
9-10   &  (5,5)          &(3,4,4,4,4,6;6)       &  (6,9)                      &(9,5) & (4,6,6,6,6,6;6)                               &                                                     &  (12,12,11,13,13;5)            \\
 &               & (3,5,5,4,4,6;6)         &   (11,6)                   &         &   (6,6,6,8,8,8;6)                            &                                                      &(15,15,13,13,13;5)      \\
        &                  &(3,4,4,4,6,6;6)        &  (17,6)                    &             &(8,8,10,10,10,10,10;7)               &                                                     &                                                \\    
               &                  &(5,4;5,6)              &                         &             &                                             &                                                     &                                                \\  \hline   
11      &(5,5)            &(3,4,4,4,4,6;6)        & (11,6)                      &              & (4,6,6,6,6,6;6)                           &                                                     & (12,12,11,13,13;5)            \\
       &                         &(3,5,5,4,4,6;6)        & (17,6)                     &               & (6,6,6,8,8,8;6)                           &                                                    & (15,15,13,13,13;5)          \\
    &                          &(3,4,4,4,6,6;6)        &                               &              &  (8,8,10,10,10,10,10;7)            &                                                    &                                         \\
        &                        &(5,4;5,6)          &                               &              &                                                    &                                                    &                                          \\\hline   
12       &(5,5)  &(3,4,4,4,4,6;6)        &  (11,6)                  &               &   (4,6,6,6,6,6;6)                          &                                                  &                                               \\
      &           &(3,5,5,4,4,6;6)       &  (17,6)                    &                &  (6,6,6,8,8,8;6)                         &                                                  &                                             \\
  &           &(3,4,4,4,6,6;6)       &                                 &               &  (8,8,10,10,10,10,10;7)             &                                                  &                                             \\
      &                  &(5,4;5,6)         &                                &                &                                                    &                                                 &                                                \\
            
\end{longtable}
\end{landscape}
\vspace{4mm} \noindent 
\textbf{Acknowledgement.} This extension has been mentioned in the author's Dissertation.
%\newpage
%%%%%%%%%%
\bibliographystyle{apa}     
\bibliography{referenceJAC.bib}
%\printbibliography[heading=bibintoc,title={References}]
\end{document}